\documentclass[fleqn,usenatbib]{mnras}

\usepackage{newtxtext,newtxmath}
\usepackage{bm}

\usepackage[T1]{fontenc}
\usepackage{ae,aecompl}
\usepackage{float}

\usepackage{graphicx}	
\usepackage{amsmath}	

\newcommand{\dd}{\mathrm{d}}
\newcommand{\ee}{\mathrm{e}}
\newcommand{\ii}{\mathrm{i}}

\title[Resonances and critical-layer formation]{Tidally excited gravity waves in the cores of solar-type stars: resonances and critical-layer formation}

\author[Z. Guo et al.]{Zhao Guo$^{1}$, Gordon I. Ogilvie$^{1}$ and Adrian J. Barker$^{2}$
\\
$^{1}$ Department of Applied Mathematics and Theoretical Physics, University of Cambridge, Cambridge CB3 0WA, UK \\
$^{2}$ Department of Applied Mathematics, School of Mathematics, University of Leeds, Leeds LS2 9JT, UK \\
}

\date{Accepted XXX. Received YYY; in original form ZZZ}

\pubyear{2022}

\begin{document}
\label{firstpage}
\pagerange{\pageref{firstpage}--\pageref{lastpage}}
\maketitle

\begin{abstract}
We simulate the  propagation and dissipation of tidally induced nonlinear gravity waves in the cores of solar-type stars. We perform hydrodynamical simulations of a previously developed Boussinesq model using a spectral-element code to study the stellar core as a wave cavity that is periodically forced at the outer boundary with a given azimuthal wavenumber and an adjustable frequency. For low-amplitude forcing, the system exhibits resonances with standing g-modes at particular frequencies, corresponding to a situation in which the tidal torque is highly frequency-dependent. For high-amplitude forcing, the excited waves break promptly near the centre and spin up the core so that subsequent waves are absorbed in an expanding critical layer, as found in previous work, leading to a tidal torque with a smooth frequency-dependence. For intermediate-amplitude forcing, we find that linear damping of the waves gradually spins up the core such that the resonance condition can be altered drastically. The system can evolve towards or away from g-mode resonances, depending on the difference between the forcing frequency and the closest eigenfrequency. Eventually, a critical layer forms and absorbs the incoming waves, leading to a situation similar to the high-amplitude case in which the waves break promptly. We study the dependence of this process on the forcing amplitude and frequency, as well as on the diffusion coefficients. 
We emphasize that the small Prandtl number in the centre of solar-like stars facilitates the development of a differentially rotating core owing to the nonlinear feedback of waves. Our simulations and analysis reveal that this important mechanism may drastically change the phase of gravity waves and thus the classical picture of resonance locking in solar-type stars needs to be revised.
\end{abstract} 




\section{Introduction}

Tidal dissipation in solar-type stars critically depends on the fate of the internal gravity waves generated by tidal forcing near the convective--radiative interface \citep{Zah77,Goo98,Ter98,Ogi07}. Previously, \citet[hereafter BO10]{Bar10} found that sufficiently massive exoplanets ($M \gtrsim 3M_\text{J}$ in the case of the present Sun) can induce gravity waves that become strongly nonlinear and break near the stellar centre, leading to enhanced tidal dissipation and fast orbital decay of the shortest-period exoplanets. In this scenario, which might explain the observed orbital decay of WASP-12b \citep{Wei17,Mac16, Pat20, Yee20}, an expanding central region of the star becomes synchronized with the decaying planetary orbit as it absorbs angular momentum from the incoming waves in a critical layer (CL). The study of wave breaking by BO10 using 2D simulations of the solar centre was later extended to 3D simulations \citep[hereafter B11]{Bar11} and the breaking process was analysed as an instability of the tidally induced gravity waves \citep[hereafter BO11]{Bar11a}. Tidal dissipation of gravity waves in solar-type stars has also been studied in a weakly nonlinear framework by \citet{Wei12} and \citet{Ess16}.

The observed orbital decay of some hot Jupiters motivates the study of tidal dissipation of stars at later evolutionary stages \citep{Wei17,Bai19,Yee20}. \citet{Bar20} studied tidal dissipation in stars with a wide range of masses ($0.1$--$1.6\,M_{\odot}$) as a function of age, taking into account the evolving stellar structure from the pre-main sequence to the end of the main sequence and assuming that the tidally induced internal gravity waves in radiative zones are fully damped. Recently, \citet{Ahu21} also studied the dissipation of gravity waves in the radiative zones of F, G and K-type stars from the pre-main sequence to the red giant branch.

The role of resonance locking (RL) \citep{Wit99, Wit01} in the tidal evolution of binary stars and short-period exoplanetary systems has been emphasized in recent years \citep{Bur13,Ful17}.
The frequency-dependent linear response of a star to tidal forcing typically exhibits a large number of narrow peaks corresponding to resonances with weakly damped global modes, such as the standing internal gravity waves (g~modes) in the radiative zone of a solar-type star (e.g.\ \citealt{Ogi14}, and references therein). Tidal evolution is very slow outside the resonances and rapid within them. The mode frequencies depend on the evolving stellar structure, while the forcing frequency depends on the evolving orbit and internal stellar rotation. Under certain circumstances, stellar and tidal evolution can cooperate in such a way that a resonance is entered and maintained, locking the tidal evolution to that of the stellar interior.
\citet{Ma21} concluded that RL is the dominant tidal dissipation mechanism for (i) stars with convective cores ($M \gtrsim 1.1M_{\odot}$) and (ii) stars with radiative cores and less-massive planets ($M \lesssim 0.3 M_\text{J}$). For massive planets ($M \gtrsim 3M_\text{J}$) and intermediate-mass planets ($0.3M_\text{J} \lesssim M \lesssim 3M_\text{J}$), \citet{Ma21} favoured the wave breaking mechanism (BO10) and the weakly nonlinear mode-coupling mechanism \citep{Ess16}, respectively, for tidal dissipation. \citet{Zan21} focused on the orbital circularization of stellar binaries with masses from one to two solar masses via RL. They found that the circularization process involving RL with $m=0$ g~modes occurs primarily in the pre-main-sequence phase. A major concern with the applicability of RL is that the resonantly amplified waves may break, limiting the tidal torque that can be achieved. When these waves are damped, they are also likely to cause the star to rotate differentially, as was found in simulations of the core of a solar-type star by BO10 and hypothesized at the surface of an early-type star by \citet{Gol89}.

BO10 (in their Section 10.2; see also \citealt{Bar11b}) discussed the alternative possibility that, even if the internal gravity waves generated by tidal forcing do not exceed the critical amplitude needed to break near the centre of a solar-type star, the weak damping of the subcritical waves by radiative diffusion would eventually deposit enough angular momentum to spin up the central region of the star and generate a critical layer. This would absorb subsequent waves, and lead to a similar outcome to that in which the waves break promptly.

In this paper, we carry out simulations and analysis to explore this scenario in detail. 
As we will show in Section~\ref{s:simulations}, we can only simulate a parameter regime with much larger viscosity
than a realistic stellar interior. Nevertheless, we believe that the simulations are still illuminating and can provide some clues as to the physical processes at work. We aim to explain the processes semi-analytically and thereby extrapolate the results to realistic stars in which the viscosity is very small. We also make connections with the important problem of tidal resonances described above. We reveal an important aspect of the problem, not considered by BO10, which is that a partial spin-up of the core due to wave damping can significantly alter the phase of the waves and therefore drastically affect the conditions for resonance. We argue that this mechanism needs to be taken into account in future studies of both tidal resonances and wave breaking in solar-type stars.

The remainder of this paper is structured as follows. In Section~\ref{s:model}, we introduce the Boussinesq model of the central region and give the conservative forms of the equations for angular momentum and entropy. In Section~\ref{s:simulations}, we present details of our numerical methods and describe the results of the simulations.
In Section~\ref{s:analytical}, we implement an analytical approach to study the linear waves and their interactions with the slowly evolving mean flow.

Finally, in Sections~\ref{s:implications} and~\ref{s:conclusions}, we further discuss the astrophysical implications of our results and draw conclusions.

\section{Boussinesq model of the central region}
\label{s:model}

\subsection{Physical properties}
\label{s:physical}

The density profile of the solar interior shows a plateau near the centre, justifying the Boussinesq approximation adopted in BO10 for sufficiently short wavelengths. We have tabulated the basic physical parameters in the solar centre in Table 1. Note that the small Prandtl number $\text{Pr}=\nu/\kappa$ indicates that thermal diffusion (with diffusivity $\kappa$) is much more important than viscous dissipation (with kinematic viscosity $\nu$). For values of $\nu$ and $\kappa$ through the whole solar interior, see \citet{Cal16}.

In general, both thermal and compositional gradients contribute to buoyancy forces in stars and different diffusivities apply to each component. In the core of the Sun, the squared Brunt-V\"{a}is\"{a}l\"{a} frequency $N^2$ is primarily due to a compositional gradient; the thermal gradient contributes only about 10\% to the total buoyancy. 
As described below, our simulations include only one form of buoyancy. We defer to Section~\ref{s:implications} a discussion of multiple buoyancy effects with applications to stars. \citet[][in their Section~4]{Gar15} 
provide some evaluations of the Prandtl number and the ratio of compositional and thermal diffusivities in various stars and for various evolutionary stages.

\begin{table}
	\centering
	\caption{Properties of the solar core.}
	\label{tab:example_table}
	\begin{tabular}{lccc} 
		\hline
		Parameters & Symbol & Values & Unit\\
		\hline
		Density & $\rho$  & $152$ & g cm$^{-3}$ \\
		Temperature & $T$ & $1.57 \times 10^{7}$ & $K$ \\
		Buoyancy gradient & $C=\dd N/\dd r$ & $8 \times 10^{-13}$ & cm$^{-1}$ s$^{-1}$ \\
		Thermal diffusivity & $\kappa$ & $1.62\times 10^{5}$ & cm$^{2}$ s$^{-1}$\\
		Kinematic viscosity & $\nu$ & $3.8$ & cm$^{2}$ s$^{-1}$\\
		\hline
		Prandtl number & $\text{Pr}=\nu/\kappa$ & $2 \times 10^{-5}$ &  \\
		\hline
	\end{tabular}
\end{table}

\subsection{Basic equations}

Following BO10, the basic equations relevant for describing gravity waves and mean flows near the centre of a solar-type star are given in vector form by
\begin{align}
&D\bm{u}=-\bm{\nabla} q+\bm{r}b+\nu\nabla^2 \bm{u},\label{du}\\ 
&Db+C^2 \bm{r \cdot u} =\kappa \nabla^2 b,\label{db}\\
&\bm{ \nabla \cdot u}=0,\label{divu}
\end{align}
where $D=\partial_t + \bm{u \cdot \nabla}$ is the Lagrangian time-derivative following the fluid velocity $\bm{u}$, $q$ is a modified pressure perturbation and $b$ is a buoyancy variable (proportional to the density perturbation), while $\nu$ (kinematic viscosity) and $\kappa$ (thermal diffusivity) are constants. The constant $C$ measures the stable stratification in the stellar core; the buoyancy frequency of the basic state is represented near the centre of the star as $N=Cr$, which is appropriate for solar-type stars.

This model is a variant of the Boussinesq equations \citep{Spi60} adapted to the geometry and conditions near the centre of a solar-type star. It was derived formally by asymptotic analysis in the ideal case by BO10; here we include viscosity and thermal diffusion acting on the perturbation variables (as in BO11).
The model is nonlinear because of the $\bm{u \cdot \nabla u}$ and $\bm{u \cdot \nabla} b$ terms. The essential approximations involved are that (i) the region under consideration is close to the centre of the star, where the density is nearly constant and the gravitational acceleration is nearly proportional to the radius; (ii) velocities and rates of change are highly subsonic, allowing the elimination of sound waves and implying a non-divergent velocity field; (iii) the timescales involved are short compared to the nuclear timescale on which the stellar structure evolves; (iv) magnetic fields are dynamically unimportant. The model allows an accurate description of the inner wavelengths of low-frequency (high radial order) g~modes in solar-type stars and the associated nonlinear processes, while eliminating unimportant acoustic effects.

The Boussinesq model can be derived in 3D, for a spherical star. However, the equivalent 2D model, being a cylindrical representation of the central region of a star, is a useful reduced model for both analytical work and numerical simulations (BO10). Previous work has shown that the 2D and 3D models behave in qualitatively similar ways (BO10; B11).

As mentioned before, a limitation of the current model is that only a single contribution to the buoyancy is included, which we here identify as a thermal (entropy or temperature) perturbation on which a thermal diffusivity $\kappa$ acts.

\subsection{Conservation of angular momentum and buoyancy}
\label{s:conservation}

Working in 2D and in polar coordinates $(r,\phi)$,
the equation for the conservation of angular momentum (AM) is
\begin{align}
&\frac{\partial}{\partial t}(ru_{\phi}) + \frac{1}{r}\frac{\partial}{\partial r} \left[r^2u_r u_{\phi} -\nu r^3 \frac{\partial}{\partial r} \left(\frac{u_{\phi}}{r} \right)  \right]\nonumber\\
&\qquad\qquad+\frac{1}{r}\frac{\partial}{\partial \phi} \left(r u^2_{\phi} +r q - \nu \frac{\partial u_{\phi}}{\partial \phi} -2 \nu u_r \right) =0 .\label{am}
\end{align}
After taking an azimuthal average (denoted by an overbar), we have

\begin{equation}
\frac{\partial}{\partial t}(r^2\overline{\Omega}) + \frac{1}{r}\frac{\partial}{\partial r} \left(r^2 \overline{u_r u_\phi} -\nu r^3 \frac{\partial\overline{\Omega}}{\partial r}  \right) =0 ,
\end{equation}
where $\Omega=u_\phi/r$ is the angular velocity.
The two terms in the second pair of brackets correspond to the advective and viscous AM fluxes per radian, respectively. The advective flux is associated with the $r\phi$ component of the Reynolds stress, i.e.\ the correlation of velocity fluctuations.



The conservative form for buoyancy is
\begin{equation}
\frac{\partial b}{\partial t} +\frac{1}{r} \frac{\partial}{\partial r} \left(ru_rB - \kappa \frac{\partial b}{\partial r} \right) + \frac{1}{r} \frac{\partial}{\partial \phi} \left(u_{\phi}B - \frac{\kappa}{r} \frac{\partial b}{\partial \phi} \right) =0,
\end{equation}
where $B=b+\frac{1}{2}C^2r^2+\text{constant}$ is the total buoyancy, including the stable stratification of the basic state.
Azimuthal averaging leads to
\begin{equation}
\frac{\partial\overline{b}}{\partial t} +\frac{1}{r} \frac{\partial}{\partial r} \left(r\overline{u_rb} - \kappa \frac{\partial\overline{b}}{\partial r} \right)=0,
\end{equation}
which shows radial fluxes of buoyancy due to advection (or correlation of fluctuations) and diffusion.

If the buoyancy is of thermal or compositional origin, respectively, then the conservation of $B$ can be understood as that of entropy or substance. Note that the production of entropy by viscous heating is negligible in the Boussinesq approximation. The reason that the diffusivity $\kappa$ appears to act only on the buoyancy perturbation $b$ is that the diffusive term $\kappa\nabla^2\left(\frac{1}{2}C^2r^2\right)$ is balanced in the basic state by the production of entropy or heavy elements due to nuclear reactions.

\subsection{Gravity waves in the Boussinesq model}
\label{s:gravity}

As shown by BO10,
in the absence of diffusion, the Boussinesq model admits free oscillation modes in the form of standing internal gravity waves (g~modes). A wave of this type depends on $\phi$ and $t$ through a phase factor $\exp(\ii m\phi-\ii\omega t)$, where $m$ is the azimuthal wavenumber (a non-zero integer) and $\omega$ is the angular frequency (a real number). The wave is therefore stationary in a frame that rotates with the angular pattern speed $\Omega_\text{p}=\omega/m$.

The radial structure of these g~modes is determined by Bessel's equation of order $m$. The wave solution that is regular at $r=0$ has $ru_r\propto b\propto J_m(kr)$, where $J$ denotes the Bessel function of the first kind and $k=C/\Omega_\text{p}$ is the radial 
wavenumber. In an unbounded system, $k$ can take any positive value and there is a continuous spectrum of g~modes; this is because, in our model that is designed to describe the central region of the star, $N^2$ increases without bound as $r\to\infty$.
When our model is considered in a finite domain with an outer boundary, the g~modes become discrete; the allowed values of $k$ (and therefore of $\omega$) are determined by the outer boundary condition. In particular, a rigid outer boundary at $r=R$ imposes the condition $J_m(kR)=0$, which has discrete solutions, e.g.\ $kR=5.136,8.417,11.620,14.796,17.960$, etc., in the case $m=2$ that is most relevant for tidally forced waves.

A remarkable property of the 2D system is that a linear wave solution is valid for any amplitude because the nonlinear terms $\bm{u}\bm{\cdot}\bm{\nabla}\bm{u}$ and $\bm{u}\bm{\cdot}\bm{\nabla}b$ are exactly zero (BO10). However, if the amplitude is sufficiently large, the wave overturns the stratification and may be expected to break because of instability (BO11). A detailed form of the wave solution in the case $m=2$ is

\begin{align}
  &u_r=\frac{\Omega_\text{p}^2}{C}\,\text{Re}\left[8A\,\frac{J_2(x)}{x}\,\ee^{2\ii\tilde\phi}\right],\label{ur}\\
  &u_\phi=\frac{\Omega_\text{p}^2}{C}\,\text{Re}\left[4\ii A\,J_2'(x)\,\ee^{2\ii\tilde\phi}\right],\label{up}\\
  &q=\frac{\Omega_\text{p}^4}{C^2}\,\text{Re}\left[4\ii A\,xJ_2'(x)\,\ee^{2\ii\tilde\phi}\right],\\
  &b=\Omega_\text{p}^2\,\text{Re}\left[-4\ii A\,J_2(x)\,\ee^{2\ii\tilde\phi}\right],\label{q}
\end{align}
where $A$ is a dimensionless complex amplitude, $x=kr$ is a dimensionless radial coordinate and $\tilde\phi=\phi-\Omega_\text{p}t$ is the azimuthal angle in a frame rotating with the wave.
The overturning condition (for the squared buoyancy frequency to be negative) $C^2r^2+r\frac{\partial b}{\partial r}<0$ is satisfied at some point if $|A|>1$;  this is equivalent to the condition that the angular velocity perturbation $u_\phi/r$ exceeds the angular pattern speed $\Omega_\text{p}$ at some point (at least in two dimensions; BO10). Note that the maximum value of $4J_2'(x)/x$ is $1$ and occurs in the limit $x\to0$, so the wave is most likely to break at the centre of the star.

If the outer boundary condition specifies the radial velocity to be
\begin{equation}
  u_r=U\cos(m\phi-\omega t)\qquad\text{at}\quad r=R,
\end{equation}
where $U$ is a constant, then the forced wave solution  (in the case $m=2$, so that $\omega=2\Omega_\text{p}$) is
given by equations (\ref{ur})--(\ref{q}), with dimensionless amplitude
\begin{equation}
  A=\frac{UC}{8\Omega_\text{p}^2}\frac{X}{J_2(X)},\label{a}
\end{equation}
where $X=kR$ is the value of $x$ at the outer boundary.
A compatible boundary condition for the buoyancy is
\begin{equation}
  b=\frac{UC^2R}{\omega}\sin(m\phi-\omega t).
\end{equation}
The amplitude $A$ diverges when the forcing frequency matches the frequency of a g~mode, i.e.\ when $J_2(X)=0$, corresponding to a resonance.

Viscosity and thermal diffusion cause these modes to be damped and their resonances to be moderated. The linear theory is developed in detail in Section~\ref{s:analytical} below, including the effects of changes to the background state.

\begin{figure*}
    \centering
	\includegraphics[width=1.2\columnwidth,angle=90]{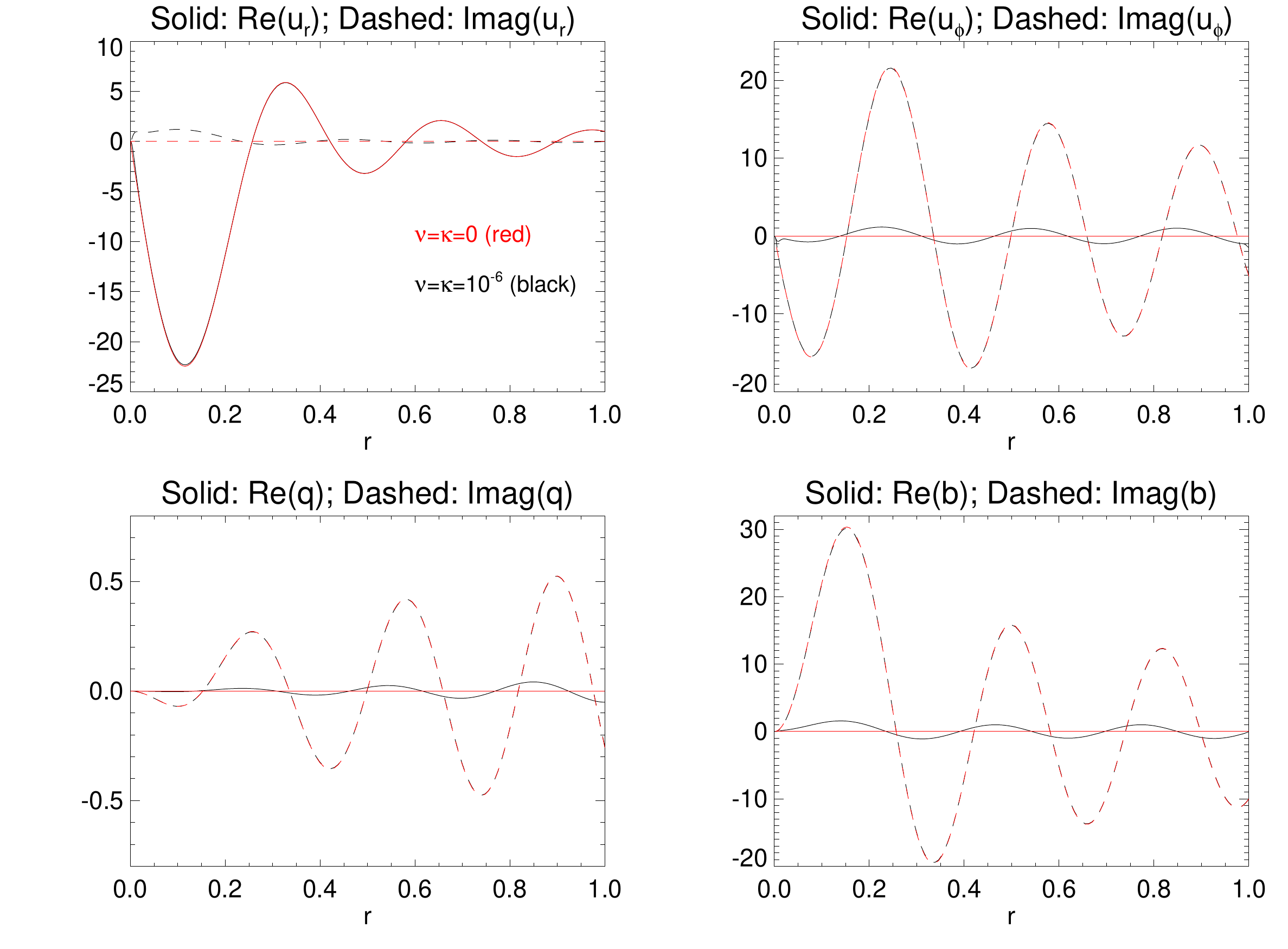}
    \caption{Linear wave solutions for $u_r$, $u_{\phi}$ (radial and azimuthal velocities), $q$ (pressure) and $b$ (buoyancy). The (arbitrary) forcing amplitude and frequency are set to $U=1$ and $\omega=0.1$, and units are adopted such that $R=1$ and $C=1$. Two cases with $\nu=\kappa=10^{-6}$ (black) and $\nu=\kappa=0$ (red) are presented; the real parts are shown as solid lines, the imaginary parts as dashed lines.}
    \label{fig:linsol}
\end{figure*}

In Fig.~\ref{fig:linsol}, we show the linear wave solution (see Section~\ref{s:analytical}) for a particular forcing frequency, without viscosity and thermal diffusion, in red. This is in good agreement with equations (\ref{ur})--(\ref{q}) (omitting the time dependence). Note that $u_{\phi}$, $q$ and $b$ are in phase with each other, while $u_r$ is $\pi/2$ out of phase. The real parts of $u_{\phi}$, $q$ and $b$ and the imaginary parts of $u_r$ are strictly zero for the ideal wave solutions. Curves in black correspond to wave solutions with $\nu=\kappa=10^{-6}$ (in units of $CR^3$); the damping significantly affects the imaginary part of $u_r$ and the real parts of $u_{\phi}$, $q$ and $b$.

The solution given in equations (\ref{ur})--(\ref{q}) and (\ref{a}) assumes that the internal gravity wave reflects perfectly from the centre of the star (or, more accurately, from a turning point near the centre). The AM flux in the resulting standing wave is zero because the radial and azimuthal velocity perturbations are out of phase. In the opposite regime in which the wave is perfectly absorbed in the central region, the appropriate solution is an inwardly travelling wave, in which the (real) Bessel function $J_2$ is replaced by the (complex) Hankel function\footnote{The Hankel function of the first kind is defined in terms of the Bessel functions of the first and second kinds by $H_2^{(1)}(x)=J_2(x)+\ii Y_2(x)$. The corresponding internal gravity wave has a radially outward phase velocity but a radially inward group velocity.} $H_2^{(1)}$. This solution diverges at $r=0$, but the wave is supposed to have been absorbed before reaching the centre. The travelling wave carries an inward AM flux [see equations~(\ref{am}), (\ref{ur}) and (\ref{up})] \begin{equation}
  T_\text{tw}=-\int_0^{2\pi}\rho r^2u_ru_\phi\,\dd\phi= \frac{\rho R^2|U|^2}{\left|H_2^{(1)}(X)\right|^2}\approx\frac{\pi X}{2}\rho R^2|U|^2,\label{ttw}
\end{equation}
where\footnote{The second equality in equation~(\ref{ttw}) uses the Wronskian property $J_2(x)Y_2'(x)-Y_2(x)J_2'(x)=2/(\pi x)$.} the approximation is valid for $X\gg1$. For a given forcing amplitude $U$, then, the `tidal' torque acting on the system in the travelling-wave regime is inversely proportional to the forcing frequency $\omega$, because $X=kR$ and $k=C/\Omega_\text{p}=Cm/\omega$.

Using the analysis in Section~\ref{s:analytical}, it can be shown that, in the standing-wave regime when diffusion is weak, the torque can be written as
\begin{equation}
  T_\text{sw}=\frac{\pi X}{2}\rho R^2|U|^2\tau(X),
\label{tsw}
\end{equation}
where
\begin{align}
  \tau(X)&=\epsilon\frac{J_2(X)^2-J_1(X)J_3(X)}{|J_2(X-\ii\epsilon)|^2}\\
  &\approx\frac{\epsilon}{\sin^2\left(X+\frac{\pi}{4}\right)+\epsilon^2}\label{tau_approx}
\end{align}
is a positive dimensionless function
and
\begin{equation}
  \epsilon=\frac{(\nu+\kappa)k^2}{2\omega}X
\end{equation}
is a small positive dimensionless quantity that measures the damping of the resonances. It can be interpreted as a measure of the ratio of the radial group travel time $R/(\omega/k)$ (for a low-frequency gravity wave) to the combined viscous and thermal damping time $2/((\nu+\kappa)k^2)$. For the parameters used in the non-ideal case in Fig.~\ref{fig:linsol}, $\epsilon=0.08$. The approximation (\ref{tau_approx}) is valid for $X\gg1$ and $\epsilon\ll1$; it gives a mean value $\approx1$ (so that the frequency-averaged standing-wave torque agrees with the travelling-wave torque), but exhibits resonant peaks of height $1/\epsilon$ separated by troughs of height $\approx\epsilon$.

\section{Two-dimensional numerical simulations}
\label{s:simulations}

\subsection{Set-up and basic parameters}
We use the spectral element code Nek5000 
\citep{nek5000} to solve the hydrodynamical equations (\ref{du})--(\ref{divu}). Nek5000 partitions the domain into a set of $\mathcal{E}$ non-overlapping elements, and within each element the velocity components and the pressure are represented as tensor product Legendre polynomials of order $\mathcal{N}_p-1$ and $\mathcal{N}_p-3$, respectively, defined at the Gauss-Lobatto-Legendre and Gauss-Legendre points. The total number of grid points (for the flow) is $\mathcal{E}\mathcal{N}_p^3$. Spectral element methods combine the high accuracy of spectral methods, which have exponential convergence with $\mathcal{N}_p$ for smooth solutions, with the geometrical flexibility and parallel scalability of finite element methods. Nek5000 has been used widely to study problems in combustion, nuclear engineering, aerodynamics, and magnetohydrodynamics. It has also been used in astrophysics by, e.g., \citet{Bar16} for studying elliptical instability in ellipsoids with a free surface, and \citet{Fav14} for tidal flows in spherical shells.

The simulation domain is a 2D circular cavity with outer radius $R$ that is meant to represent a small fraction (typically $\approx 2$\%) of the stellar radius\footnote{In BO10, the size of simulation domain contains about 13 radial wavelengths. However, since most of the dynamics (wave breaking and critical-layer formation) we are concerned with happen in the centre, the size of the domain is not so important. A smaller number of wavelengths also allows smaller viscosities/diffusivities to be accurately simulated.}. It is designed to contain several wavelengths of the internal gravity waves of interest. We adopt $R$ as the unit of length for the numerical simulations and $(CR)^{-1}$ as the unit of time. The dimensionless parameters of the model are then the Reynolds number $\mathrm{Re}=CR^3/\nu$ and the Prandtl number $\mathrm{Pr}=\nu/\kappa$. The (thermal) P\'eclet number is $\mathrm{Pe}=CR^3/\kappa=\mathrm{Re}\,\mathrm{Pr}$.

\begin{figure}
	\includegraphics[width=\columnwidth]{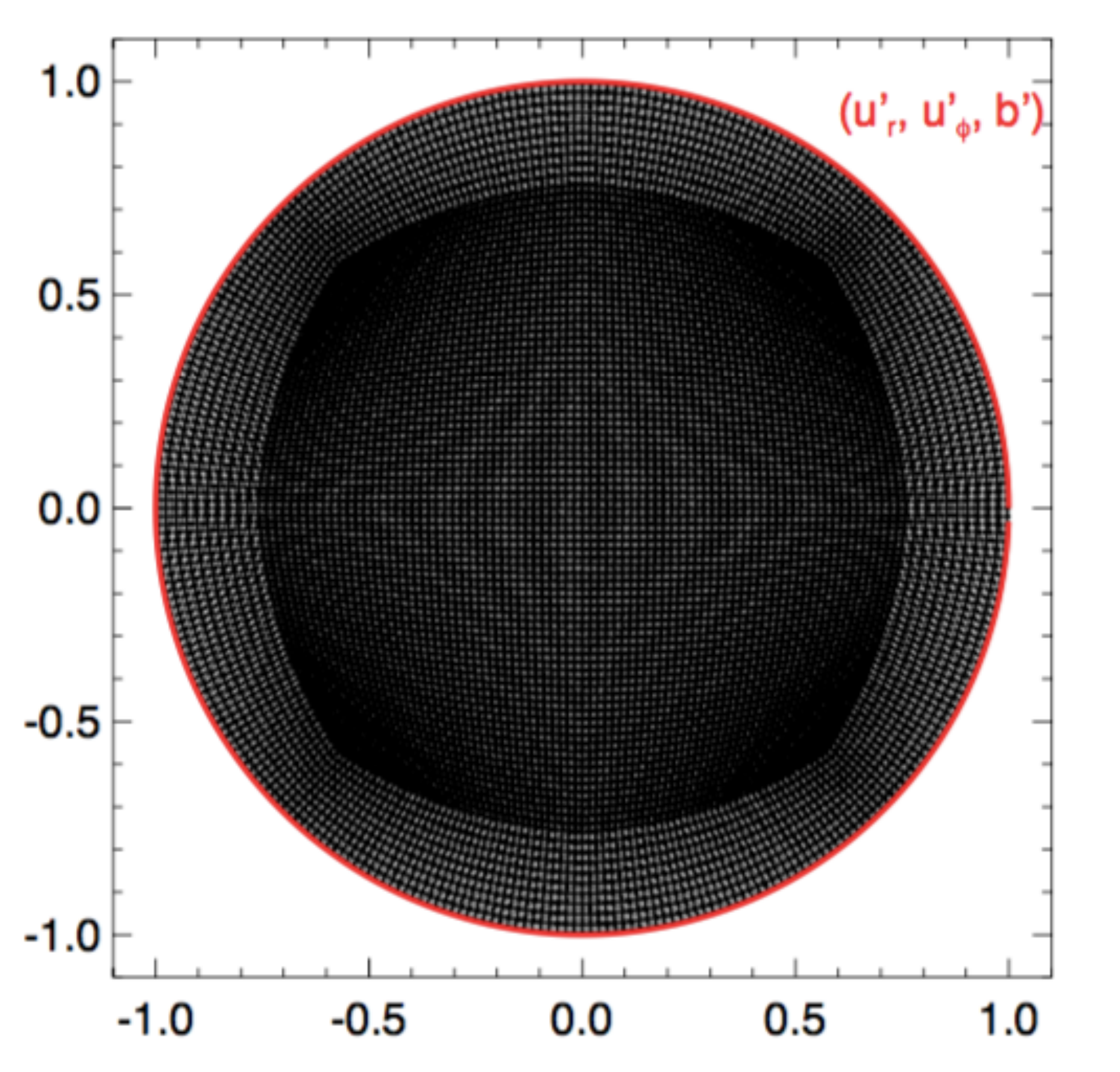}
    \caption{Illustration of the circular mesh and simulation cavity used for our simulations. This has $\mathcal{E}=9600$ elements and $\mathcal{N}_p=6$ points within each element. At the outer boundary (red circle), we force the cavity with prescribed radial and azimuthal velocities $u'_r$, $u'_{\phi}$ and buoyancy $b'$.  }
    \label{fig:mesh}
\end{figure}

We use a second-order characteristics-based time-stepping scheme for the nonlinear terms and an implicit scheme for the viscous/diffusion and pressure terms, with a variable time-step determined by a target CFL number. Our typical resolution is $\mathcal{E}=9600$ and $\mathcal{N}_p=6$ (9 for the nonlinear terms) unless otherwise specified. The nonlinear terms are fully de-aliased by using a polynomial order that is 3/2 larger for their evaluation. An example mesh and simulation domain is shown in Fig.~\ref{fig:mesh}. The mesh is quasi-Cartesian near the centre, to avoid a coordinate singularity at $r=0$, and consists of circular shells (to double precision) in the outer portions.

For our initial conditions, we adopt zero velocity ($\boldsymbol{u}=\boldsymbol{0}$) and buoyancy perturbation ($b=0$). To aid initialisation of instability though, we introduce random noise of amplitude $
0.5\times 10^{-3}$ to the buoyancy variable in the inner $50\%$ of the domain (and omitting the inner $2\%$). We force gravity waves by adopting an outer boundary condition at $r=1$ of 
\begin{eqnarray}
u_r=U \cos(m\phi-\omega t), \;\;\; u_{\phi}=0, \;\;\; b=(U/\omega)\sin(m\phi-\omega t),
\label{bcs_simulation}
\end{eqnarray}
which is designed to match the behaviour expected for the linear gravity-wave solution (Section~\ref{s:gravity}). We focus on $m=2$ as this is usually the most important component of the tidal response, but we vary the tidal forcing frequency $\omega$ and forcing amplitude $U$.


The physical parameters of our simulations include the forcing frequency $\omega$, the forcing amplitude $U$, the kinematic viscosity $\nu$ and the thermal diffusivity $\kappa$. The parameter values that we adopt are listed in Table~\ref{tab:example_table}. Note that in the simulations, we manage to reach $\mathrm{Re} \sim 10^5-10^6$ (where\footnote{An alternative definition of $\mathrm{Re}$ 
can be based on the properties of the simulated flow. For a typical fluid velocity $V \sim 10^{-4}-10^{-3}$, length scale $L \sim 1$ and viscosity $\nu=10^{-6}-5\times 10^{-6}$, $\mathrm{Re}=VL/\nu \sim 10^2-10^3$. } $\mathrm{Re}=CR^3/\nu=1/\nu$)  and $\mathrm{Pr}=0.1-1$. These are very different from the values in the real solar core (Section~\ref{s:physical}), as is common with many problems in astrophysical fluids. Our hope is that by understanding the physics in these simulations with accessible parameters, and by constructing analytical descriptions that can be extrapolated, we can draw astrophysically relevant conclusions.
\begin{table}
	\centering
	\caption{Summary of our simulation parameters. }
	\label{tab:example_table}
	\begin{tabular}{lccr} 
		\hline
		$\omega$ & $U$ & $\nu$ & $\kappa$\\
		\hline
		0.100 -- 0.118 & $(1,3,10) \times 10^{-5}$ & $10^{-6}$ & $10^{-6}$\\
		\hline
		0.100 -- 0.118 & $(1,3,10) \times 10^{-5}$ & $10^{-6}$ & $5 \times 10^{-6}$\\
		\hline
		0.100 & $10^{-6}$ & $10^{-6}$ & $10^{-6}$\\
		0.100 & $(1,3,4,5) \times 10^{-5}$ & $10^{-6}$ & $10^{-6}$\\
		0.100 & $10^{-4}$ & $10^{-6}$ & $10^{-6}$\\
		\hline
		0.100 & $10^{-5}$ & $10^{-6}$ & $(1,2,5) \times 10^{-6}$\\
		0.100 & $10^{-5}$ & $10^{-6}$ & $ 10^{-5}$\\
		\hline
	\end{tabular}
\end{table}

\begin{figure}
	\includegraphics[width=\columnwidth]{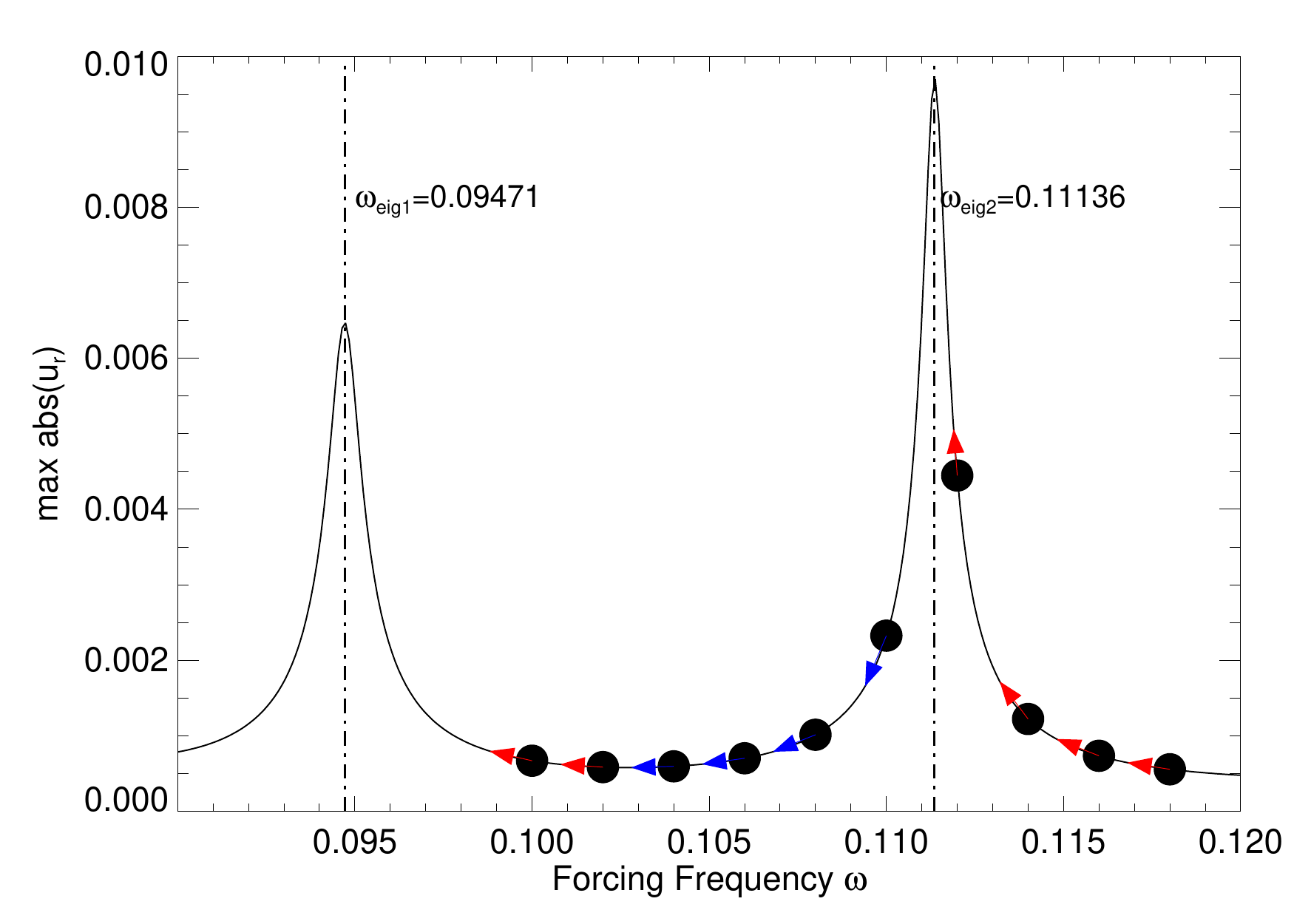}
    \caption{The linear tidal response in terms of max$(u_r)$ for low-amplitude forcing ($U=10^{-5})$ with $\nu=10^{-6}$ and $\kappa=10^{-6}$ can be represented by resonance peaks (solid lines) centred at the eigenfrequencies $\omega_{\rm{eig}1,2}$. The dots are starting frequencies adopted in our simulations. Arrows indicate the direction of the evolution as the spin-up of the central simulation domain begins. Depending on the location of the forcing frequency, the waves can either increase (red) or decrease (blue) in amplitude. }
    \label{fig:peaks}
\end{figure}

\begin{figure}
	\includegraphics[width=\columnwidth]{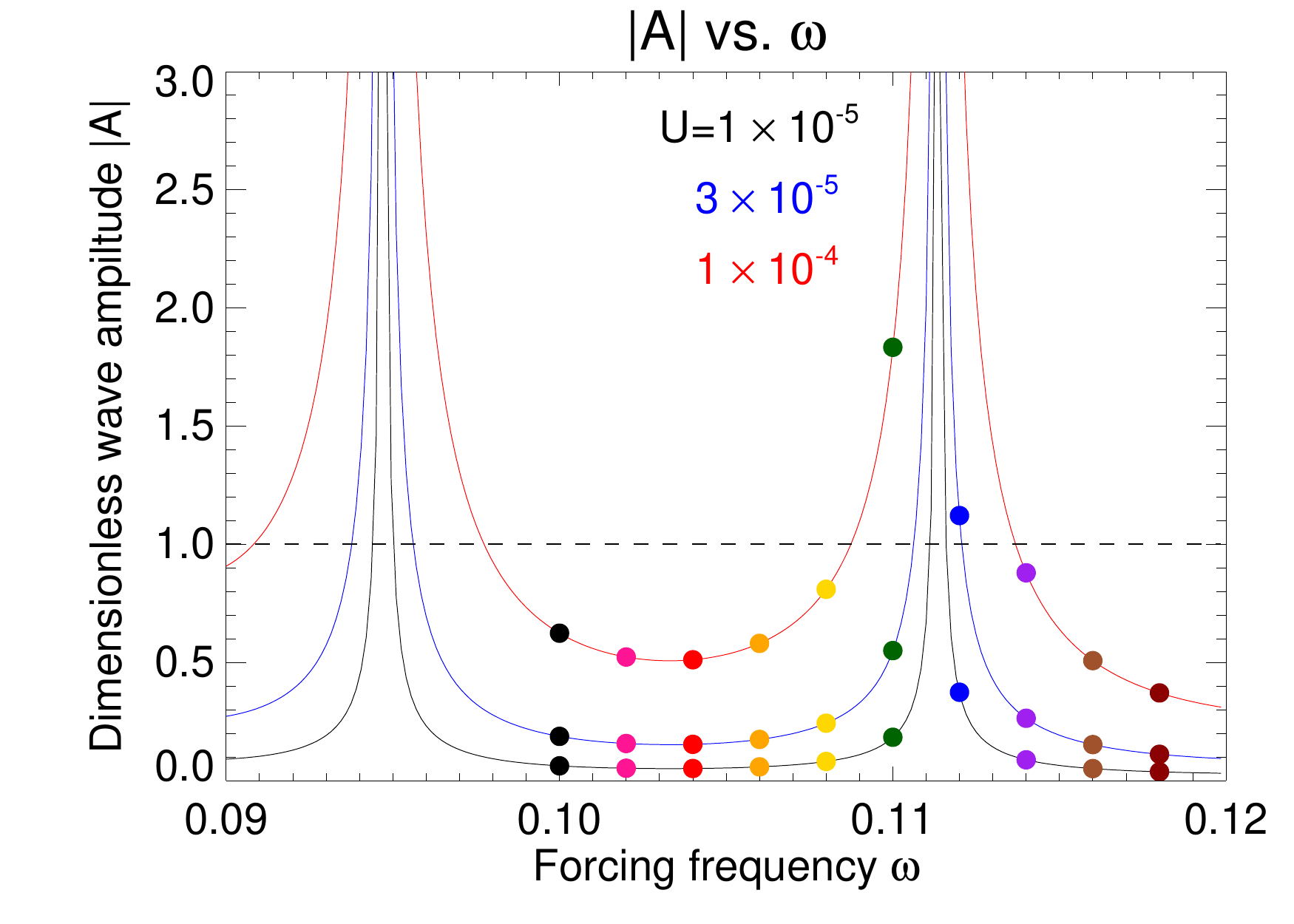}
    \caption{The dimensionless wave amplitude $|A|$ (as defined in BO11) as a function of forcing frequency $\omega$ in the standing-wave case for several forcing amplitudes $U$, with $\nu=10^{-6}$ and $\kappa=5\times 10^{-6}$. The critical amplitude for wave breaking in BO10 ($|A|=1$) is indicated by the horizontal dashed line. The forcing frequencies we scanned in the simulations are marked by the filled circles.} 
    \label{fig:A_w}
\end{figure}

\subsection{Simulation results}

The linear response to our `tidal' forcing with an amplitude of $U=10^{-5}$ is shown in Fig.~\ref{fig:peaks}, where we plot the maximum value of $|u_r|$ in the domain for the linear calculations of Section~\ref{s:gravity} (neglecting nonlinear terms). The value of $|u_r|$ is maximal in the innermost wavelength as we can see in Fig.~\ref{fig:linsol}. Fig.~\ref{fig:peaks} shows that the resonant peaks, in which the amplitude of the linear response is enhanced, are centred on the eigenfrequencies $\omega_{\rm eig}$ that exist in this frequency range. The dots represent the starting forcing frequencies that we scanned in our simulations. We will see that the evolution of the system as a result of the spin-up of the central region can be thought of as similar to a motion towards the left along this graph (even though the actual forcing frequency remains fixed).
This motion is indicated by the arrows and causes the amplitude to either increase (red) or decrease (blue). As will be shown later, the evolution away from or towards the nearest resonance peak has an important effect in determining the fate of the waves.

We present our results for different forcing amplitudes: low, intermediate, and large, corresponding to $U=1\times10^{-5}$, $3\times10^{-5}$ and $1\times10^{-4}$, respectively. The corresponding values of $A$ are shown as a function of frequency in Fig.~\ref{fig:A_w} to indicate the likelihood of wave breaking occurring.

In the figures that follow, we show the evolution of the maximum radial velocity $u_{r,\text{max}}$ of the system. Different curves indicate simulations with different forcing frequencies $\omega$. We also examine the azimuthally averaged profiles of the angular velocity and the buoyancy, and calculate the torque acting on the system.

For low-amplitude forcing ($U=10^{-5}$, Figs \ref{fig:urmax_amp1e5}--\ref{fig:2d_amp1e5}), after a transient, oscillatory phase, the system can reach a `quasi-steady' equilibrium. In essence, a linear standing wave is formed after a few wave-crossing times $t_\text{w}$. The radial group velocity of gravity waves is given by 
\begin{equation}
c_{\text{g},r}=\frac{\partial \omega}{\partial k_r} \approx \frac{\partial \omega}{\partial k} \approx \frac{\omega}{k} \approx \frac{\omega}{C/\Omega_\text{p}} =\frac{\omega^2}{m},
\end{equation} 
with $C=1$. (Here we disregard the choice of sign related to inward and outward-propagating waves.) Thus the wave crossing time (including one reflection) 
$t_\text{w} =2R/c_{\text{g},r} =2/c_{\text{g},r}=2m/\omega^2 \approx 300-400$. This is in qualitative agreement with the length of the transient phases in Fig.~\ref{fig:urmax_amp1e5}. Owing to viscous and thermal diffusion, 
the waves are slightly attenuated as they propagate across the simulation domain and AM is deposited, which slowly spins up the fluid. The central region spins up more quickly because of its much smaller moment of inertia. As shown in Fig.~\ref{fig:urmax_amp1e5}, the  $u_{r,\text{max}}$ plots are almost flat (increasing or decreasing in amplitude according to Fig.~\ref{fig:peaks}, but on a very long timescale).
However, the azimuthally averaged angular velocity, as shown in Fig.~\ref{fig:steady_amp1e5} for one of the off-resonance cases ($\omega=0.118$), is gradually increasing. In fact $\overline{\Omega}(r)$ is very slowly evolving towards a steady-state profile (dashed lines), for which we will provide an approximate theoretical expression in Section~\ref{s:equilibrium} (equation~\ref{omega_steady}). In Fig.~\ref{fig:2d_amp1e5}, we show snapshots of the radial and azimuthal velocity components $u_r$ and $u_{\phi}$ and the buoyancy variable $b$ for this simulation with low-amplitude forcing. Note that at $t=200$ the system is in the transient phase during which the waves are building up and still have a spiral form, while at later times ($t=1000$, $4000$, $8000$), the system has essentially standing waves with almost constant amplitude.

In Fig.~\ref{fig:torque_amp1e5}, we show the evolution of the specific torque $T_\text{s}$ (torque per unit mass) for the same low-amplitude forcing simulation. The torque $T_\text{s}$ is calculated as the time derivative of the specific angular momentum $dL/dt$, where $L\,\bm{e}_z=\int{\bm{r \times u} }\,\dd M/(\rho \pi R^2$)  and $\dd M =2\pi\rho r\, \dd r$ is the differential mass element (see also Section~\ref{s:evolution_torque}). The azimuthally averaged angular velocities $\overline{\Omega}$ in units of the pattern speed $\Omega_\text{p}=\omega/m$ are marked for $t=7500$ (crosses) and $t=30050$ (diamonds). Note that compared to the evolution of max($u_{r}$), it is more obvious to see the forcing-frequency dependence in the evolution of the torque plot.

For the intermediate-amplitude forcing ($U=3 \times 10^{-5}$, Figs \ref{fig:urmax_amp3e5}--\ref{fig:spinup_amp3e5}), the system undergoes a prolonged oscillatory varying phase with $u_{r,\text{max}}$ gradually increasing (evolving closer to resonance) or decreasing (evolving away from resonance), depending on whether the forcing frequency is greater or less than the nearest eigenfrequency. In the latter case the system will eventually evolve towards resonance with the eigenmode with the next-lowest frequency. In either case, the system will  eventually evolve into a resonant state in which the waves become significantly nonlinear. This process happens within the timescale of the simulation for the blue, purple, black, brown and pink lines in Fig. ~\ref{fig:urmax_amp3e5}, corresponding to $\omega=0.112$, $0.114$, $0.100$, $0.116$ and $0.102$, respectively. The central region is then spun up to an angular velocity equal to the angular pattern speed $\Omega_\text{p}=\omega/m$ of the forced waves (see Fig.~\ref{fig:spinup_amp3e5}), i.e.\ the wave frequency is zero in a frame rotating with the fluid. When this occurs, a critical layer (CL) is formed. The radial wavelength approaches zero and the wave strongly damps there, thereby depositing its angular momentum flux. A wave of infinitesimal amplitude should be strongly damped in the CL \citep{Boo67}, although nonlinear effects may allow some reflection 
(see, e.g., Section 9.2 of BO10).
The critical layer in the flow acts as an absorbing barrier for subsequent gravity waves and the system enters the travelling-wave regime.

For the high-amplitude forcing ($U=10^{-4}$, Figs \ref{fig:urmax_amp1e4}--\ref{fig:2d_amp1e4}), the system quickly undergoes wave breaking. The core is spun up to $\Omega_\text{p}$ and the critical layer gradually moves outwards as it absorbs the angular momentum transported by the incoming waves.

The wave amplitude $A$ 
is defined, as in BO10, in such a way that the wave is expected to break when $|A|>1$.
In Fig.~\ref{fig:A_w}, we showed the value of $|A|$ expected in linear theory as a function of forcing frequency $\omega$ for the three cases of low, intermediate and high-amplitude forcing.
Since the linear solution for $m=2$ (equation \ref{ur}) implies 
$u_{r,\text{max}}=0.3599|A|\omega^2$ in units such that $C=1$, wave breaking is expected for $u_{r,\text{max}}>0.3599\omega^2$.

We can compare this threshold with a simpler and more familiar estimate for wave-breaking, which is based on plane waves: $|k_r \xi_r| \gtrsim 1$, where $\xi_r$ is the radial displacement. Since $|u_r| = \omega|\xi_r|$ and $k_r \approx k = m/\omega$, this can be translated into $|u_r| \approx \omega^2/m$. For $m=2$ this yields a slightly higher threshold than the one derived above.

For typical values of $\omega$ used in our simulations, the threshold value of $u_{r,\text{max}}\ \approx 0.004$.  We can compare this threshold with the largest values of $u_{r,\text{max}}$ in Figs \ref{fig:urmax_amp3e5} and \ref{fig:urmax_amp1e4} when wave breaking/critical-layer formation occurs. For the intermediate-amplitude forcing case (Fig.~\ref{fig:urmax_amp3e5}), this maximum is roughly 0.0015--0.0025; and for the high-amplitude forcing (Fig.~\ref{fig:urmax_amp1e4}), it is about 0.004--0.005.
Thus rough agreement is observed, although the exact values may differ by a factor of a few.

In the following subsections, we give further details of the simulation results for each forcing amplitude. The diffusion coefficients are $\nu=10^{-6}$ and $\kappa=5\times 10^{-6}$ unless otherwise specified.

\subsubsection{Low-amplitude forcing ($U=10^{-5}$, for which we expect $|A|=0.04$ at $\omega=0.118$)}

\begin{figure}
	\includegraphics[width=\columnwidth]{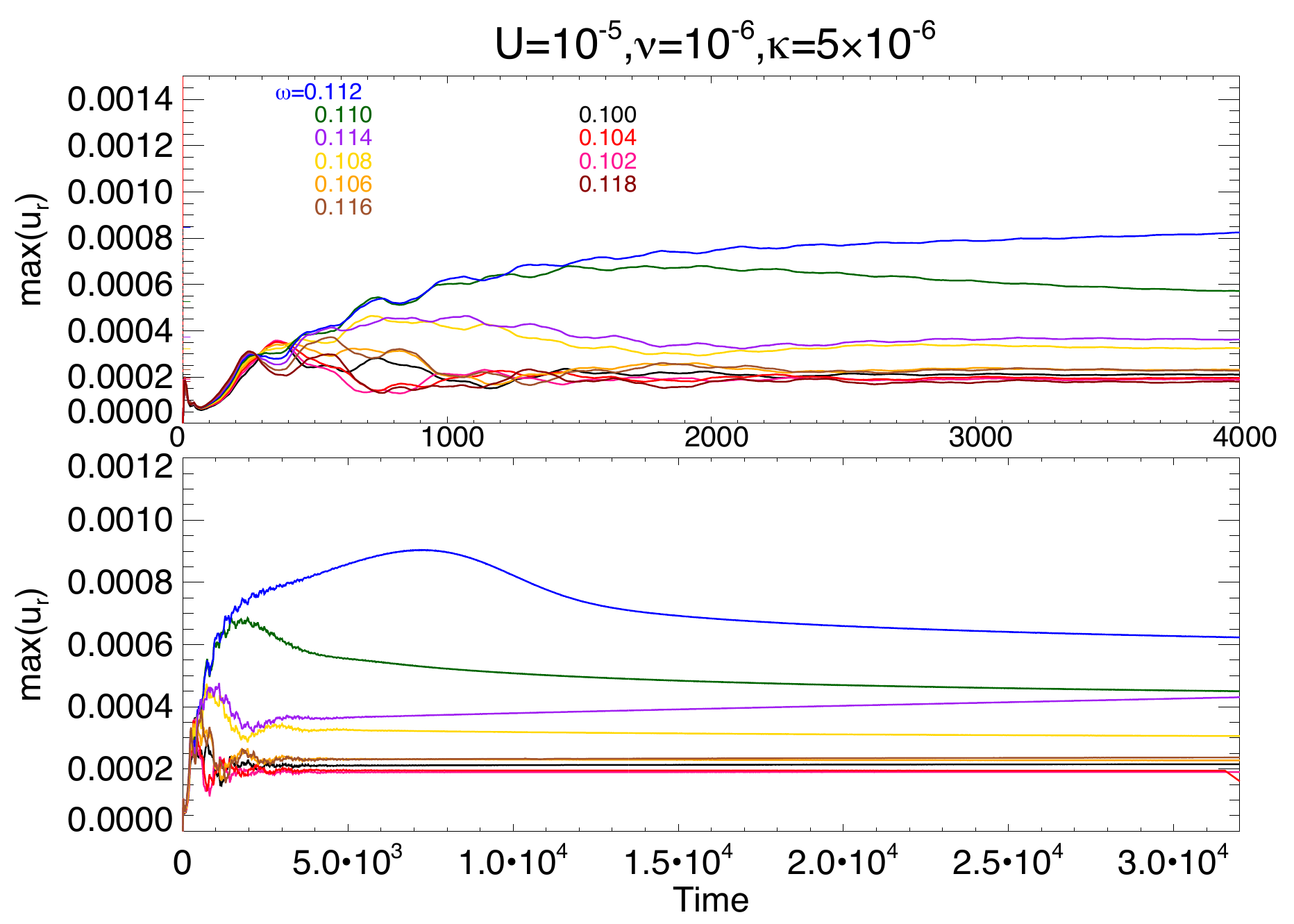}
    \caption{ $u_{r,\text{max}}$ versus time for simulations with low-amplitude forcing, $\mathrm{Pr}=0.2$, and different forcing frequencies ranging from $\omega=0.100$ to $\omega=0.118$ with a step size of $\Delta \omega=0.002$. The bottom panel shows the same cases as the top panel but for longer times. 
    }
    \label{fig:urmax_amp1e5}
\end{figure}

\begin{figure}
	\includegraphics[width=\columnwidth]{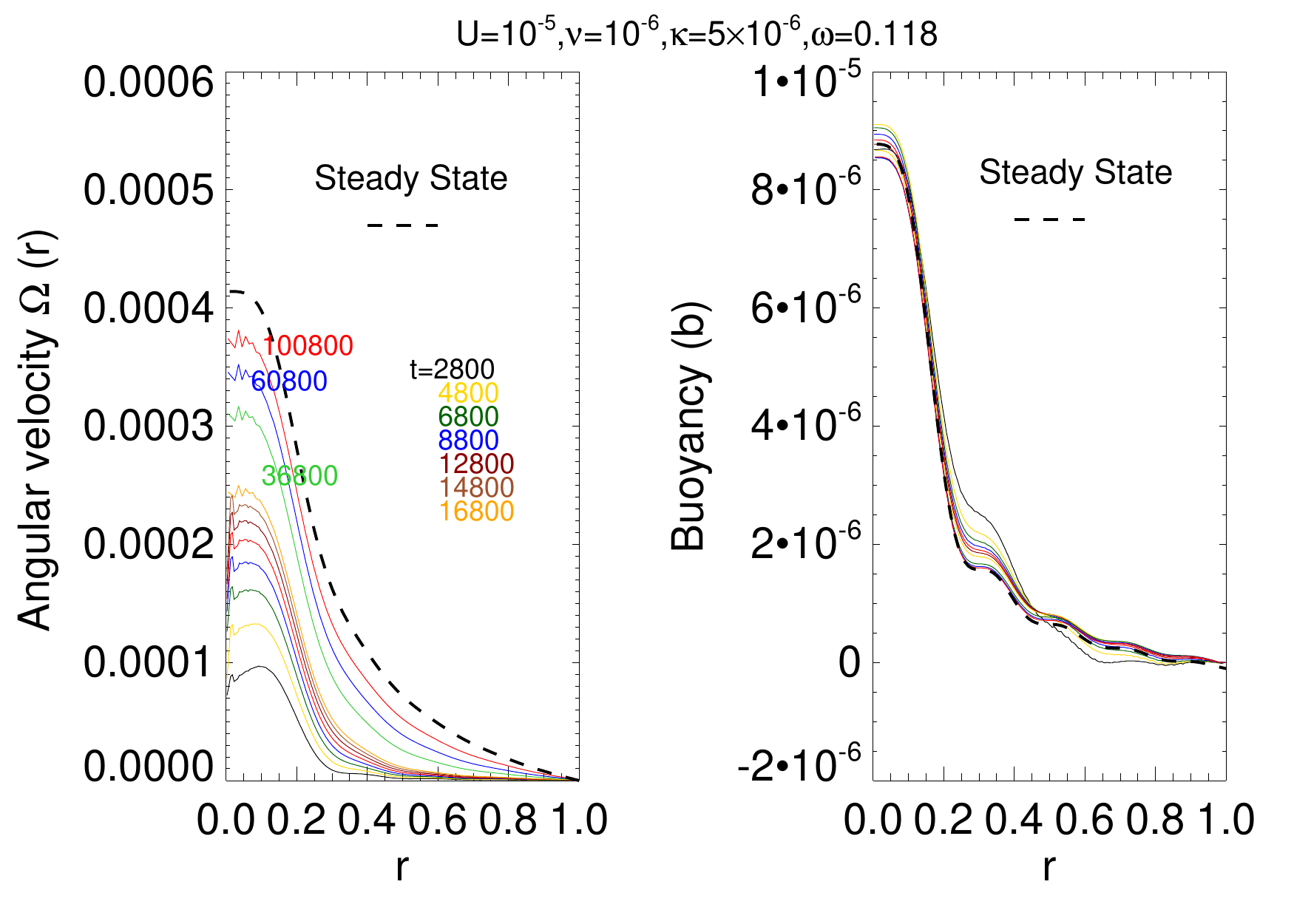}
    \caption{Azimuthally averaged profiles of angular velocity $\overline{\Omega}(r)$ and buoyancy $\overline{b}(r)$ for low-amplitude forcing with frequency $\omega=0.118$, for $\kappa= 5\times 10^{-6}$ (corresponding to the dark red line in Fig.~\ref{fig:urmax_amp1e5}) at different moments from $t=2800$ to $t=100800$. Note that $\overline{\Omega}$ remains much smaller than the angular pattern speed $\Omega_\text{p}=0.059$ throughout this simulation. The dashed lines are the steady-state predictions from Section~\ref{s:equilibrium}.}
    \label{fig:steady_amp1e5}
\end{figure}

\begin{figure}
	\includegraphics[width=\columnwidth]{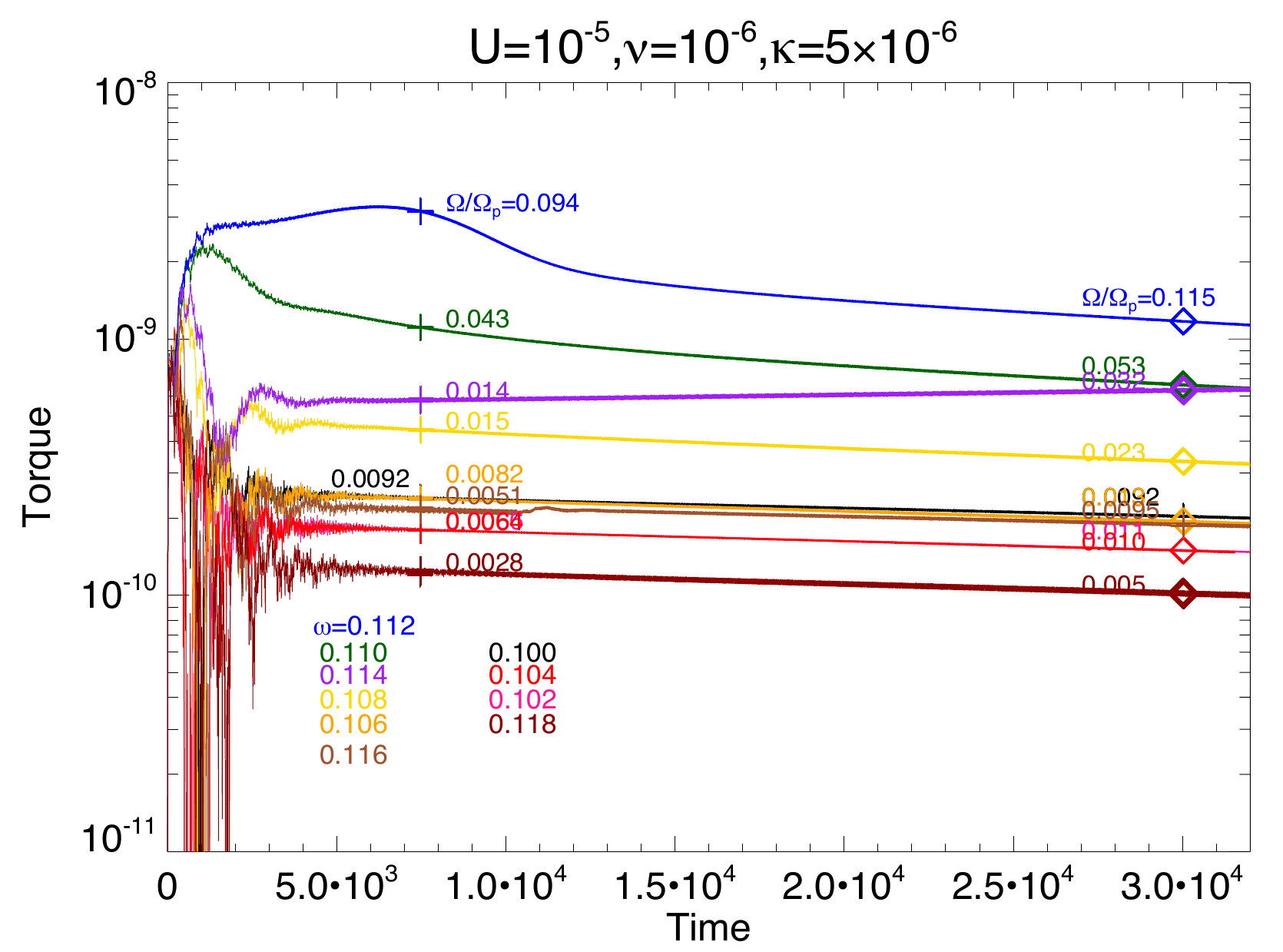}
    \caption{ Specific torque versus time for the same simulations as in Fig.~\ref{fig:urmax_amp1e5}. The azimuthally averaged angular velocities $\overline{\Omega}$ measured at the centre are labelled in units of the angular pattern speed $\Omega_\text{p}$ for $t=7500$ (crosses) and $t=30050$ (diamonds). }
    \label{fig:torque_amp1e5}
\end{figure}

\begin{figure*}
    \centering
	\includegraphics[width=1.8\columnwidth]{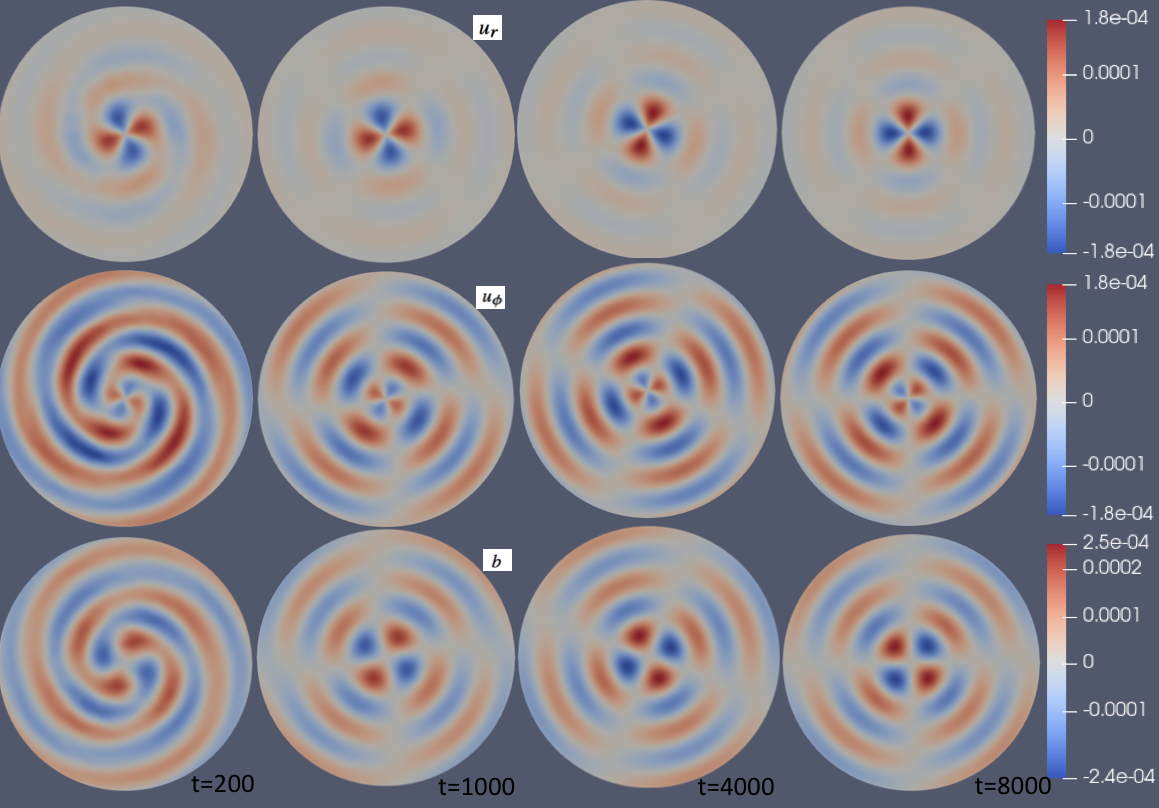}
    \caption{$u_r$ and $u_{\phi}$ (radial and azimuthal velocities)  and $b$ (buoyancy perturbation) for a simulation with low-amplitude forcing with frequency $\omega=0.118$. The diffusion coefficients are $\nu=10^{-6}$ and $\kappa=5\times 10^{-6}$.} 
    \label{fig:2d_amp1e5}
\end{figure*}

As shown in Fig.~\ref{fig:urmax_amp1e5},
the behaviour typically begins with a transient phase, in which $u_{r,\text{max}}$ undergoes an oscillatory variation with a period of $\approx 2\pi/\delta\omega$, where  $\delta\omega=\omega-\omega_\text{eig}$ is the frequency difference between the forcing frequency $\omega$ and the closest eigenfrequency ($\omega_\text{eig}=0.09471$ or $0.11136$). The smallest frequency detuning $\delta\omega$ occurs for our simulations with
$\omega=0.112$ and $0.110$ (blue and green lines, respectively), which take a very long time to settle to a `quasi-steady state'. This is also because it takes a longer time for the amplitude to build up to the larger values attained in the simulations that are closer to resonance. Note that in the case $\omega=0.112$, the system evolves even closer to resonance as a result of the gradual spin-up process and the wave amplitude increases from $t=0$ to $\sim8000$. The wave does not break, however, and the system passes through the resonance. It evolves to the left-hand side of it and the wave amplitude then begins to decrease ($t> 8000$).  For other forcing frequencies $\omega$, by the end of the simulation, $u_{r,\text{max}}$ has essentially reached a `quasi-steady state' in which the wave amplitudes are almost constant (very slowly evolving). This is typically achieved on a diffusive timescale. 
Using $k=C/\Omega_\text{p}=2/\omega$, $\nu=10^{-6}$ and $\kappa=5\times10^{-6}$, we estimate a diffusive timescale of $\sim2/(\nu+\kappa)k^2\sim 1000$, which is in agreement with that in the simulation (Fig.~\ref{fig:urmax_amp1e5}). 

The quasi-steady values of $u_{r,\text{max}}$ from a few selected simulations are overplotted in Fig.~\ref{fig:reso_peak} below (blue diamonds/crosses), where they are found to be in good agreement with the linear wave solutions (Section~\ref{s:linear_original}). 

In these simulations, the core spins up as the result of the deposition of angular momentum due to the damping of the waves by viscosity and thermal diffusion. The spin-up rate $\partial\overline{\Omega}/\partial t$ can be calculated analytically from the linear wave solutions, as long as the modification of the background remains small (see Section~\ref{s:evolution} below). 

The steady-state profiles of $\overline\Omega(r)$ and $\overline{b}(r)$, resulting from a balance between wave damping and diffusion, can also be obtained analytically (see Section~\ref{s:equilibrium} below). As can be seen in Fig.~\ref{fig:steady_amp1e5}, the steady-state solutions (equations \ref{omega_steady} and \ref{b_steady}) are in good agreement with the asymptotic behaviour of the azimuthally averaged profiles of angular velocity and buoyancy from the Nek5000 simulation for low-amplitude forcing ($U=10^{-5}$) with frequency $\omega=0.118$ ($\nu=10^{-6}, \kappa=5\times 10^{-6}$), corresponding to the dark red curve in Fig.~\ref{fig:urmax_amp1e5}. Note that, in this simulation, $u_{r,\text{max}}$ settles into a quasi-steady state after $t \sim 3000$, but the core is still gradually spinning up. From $t=4800$ to $t=14800$, the maximum angular velocity (at $r \approx 0.1$) increases from $1.3\times 10^{-4}$ to $2.3 \times 10^{-4}$. The evolution towards the equilibrium $\overline\Omega(r)$ profile is roughly exponential (we give a detailed calculation in Section~\ref{s:equilibrium} below). Note that the changes to the buoyancy profile $\overline{b}$ remain very small.

\subsubsection{Intermediate-amplitude forcing ($U=3\times 10^{-5}$, for which we expect $|A|=0.11$ at $\omega=0.118$)}

\begin{figure}
	\includegraphics[width=\columnwidth]{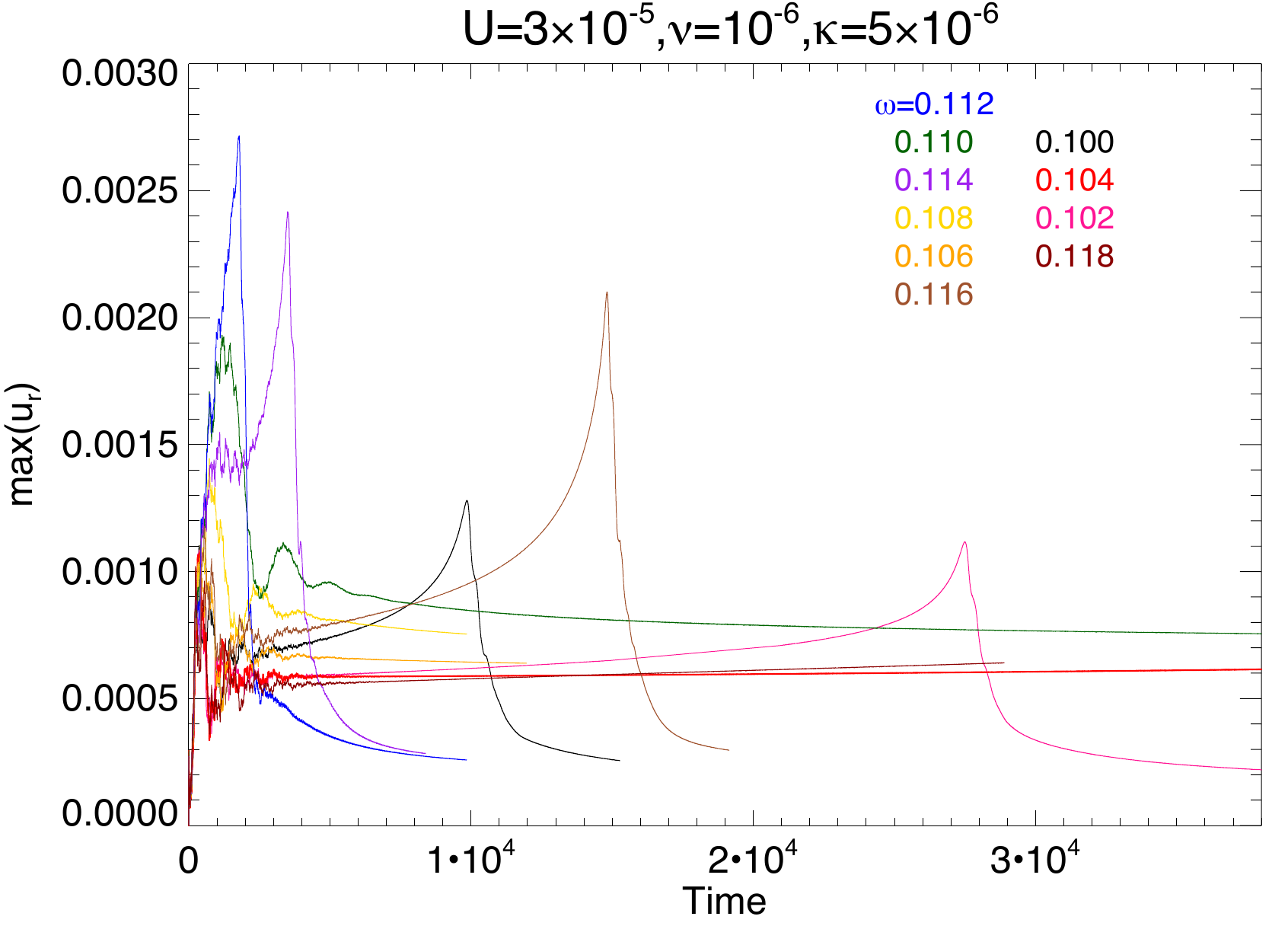}
    \caption{$u_{r,\text{max}}$ versus time for simulations with intermediate-amplitude forcing, $\textrm{Pr}=0.2$, and a series of forcing frequencies $\omega$. 
    }
    \label{fig:urmax_amp3e5}
\end{figure}

\begin{figure}
	\includegraphics[width=\columnwidth]{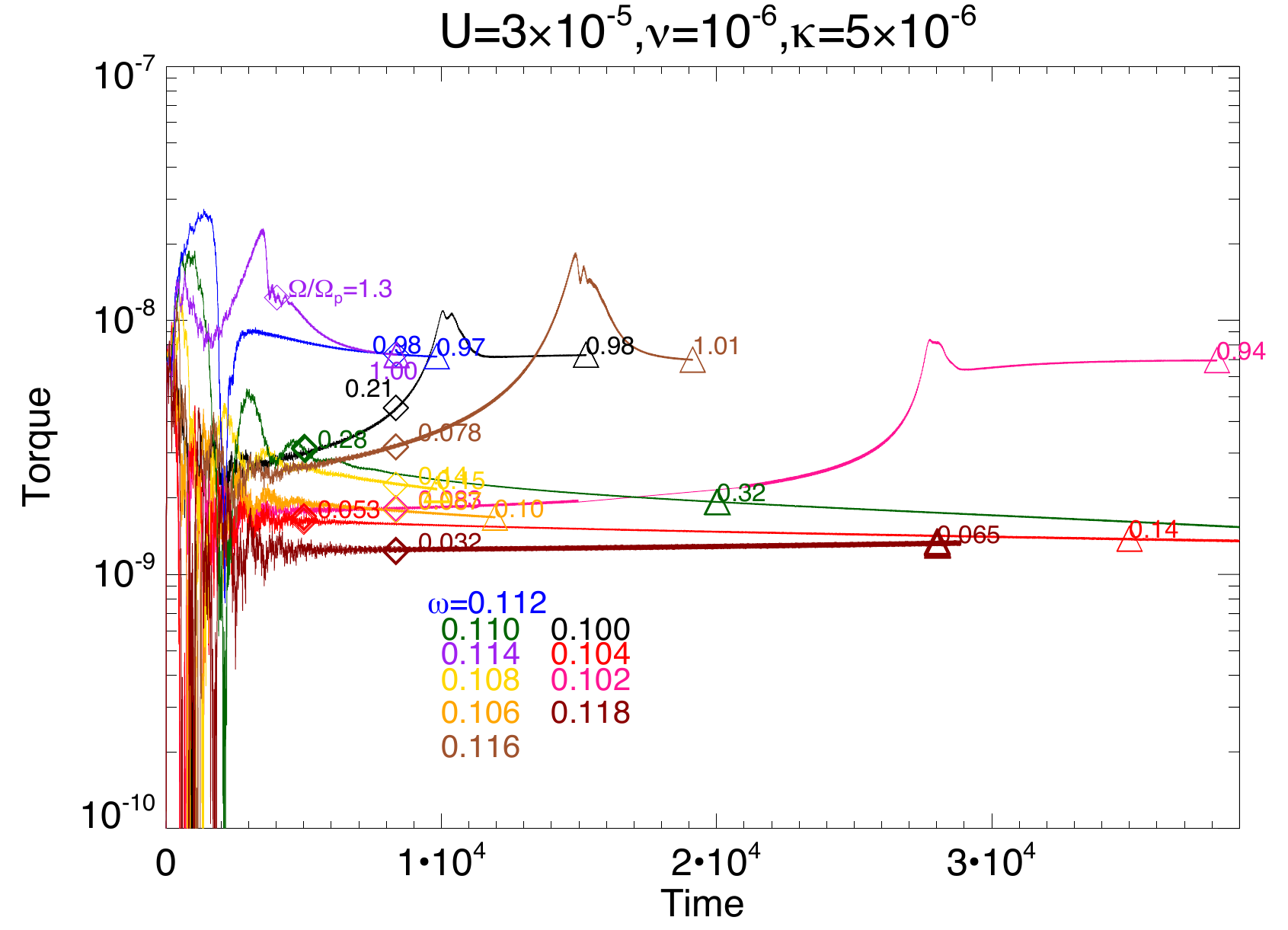}
    \caption{Specific torque versus time for the same simulations as in Fig.~\ref{fig:urmax_amp3e5}. The azimuthally averaged angular velocities $\overline{\Omega}$ measured at the centre are labelled in units of the angular pattern speed $\Omega_\text{p}$ at various times of interest, marked by the diamonds and triangles.}
    \label{fig:torque_amp3e5}
\end{figure}

\begin{figure*}
    \centering
	\includegraphics[width=1.9\columnwidth]{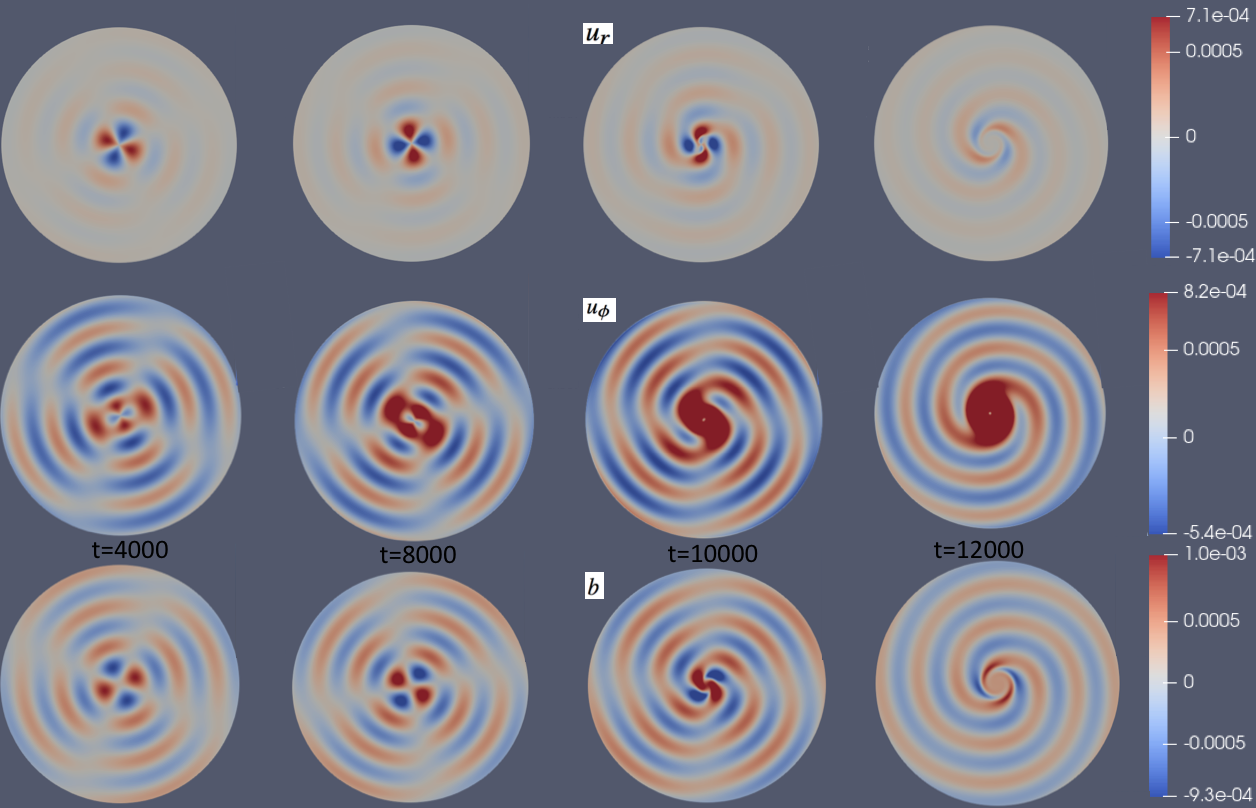}
    \caption{ $u_r$, $u_{\phi}$ (radial and azimuthal velocities) and $b$ (buoyancy) for a simulation with intermediate-amplitude forcing $U=3\times10^{-5}$ with frequency $\omega=0.100$ (corresponding to the black curve in Fig.~\ref{fig:urmax_amp3e5}). The diffusion coefficients are $\nu=10^{-6}$ and $\kappa=5\times 10^{-6}$.  } 
    \label{fig:2d_amp3e5}
\end{figure*}

Fig.~\ref{fig:urmax_amp3e5} shows the evolution of $u_{r,\text{max}}$ in simulations with intermediate-amplitude forcing. Curves with different colours represent different forcing frequencies $\omega$.

\begin{figure}
	\includegraphics[width=\columnwidth]{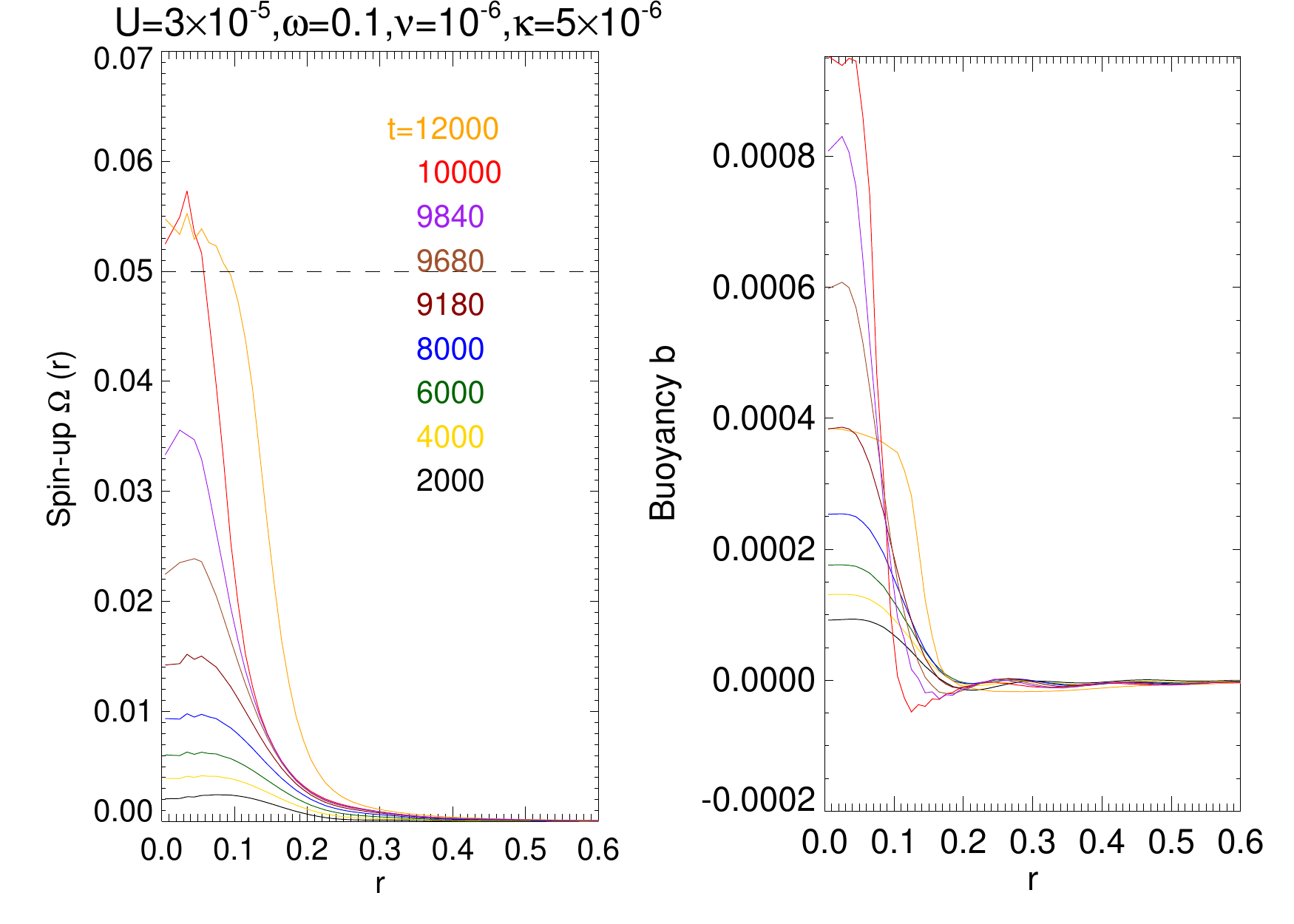}
    \caption{Azimuthally averaged profiles of angular velocity $\overline{\Omega}(r)$ and buoyancy $\overline{b}(r)$ for the case of the black curve in Fig.~\ref{fig:urmax_amp3e5} (intermediate-amplitude forcing, $\omega=0.1$, $\nu=10^{-6}$, $\kappa=5\times 10^{-6}$). A critical layer is expected to occur where $\overline{\Omega}$ matches the angular pattern speed $\Omega_\text{p}$ of the waves (dashed line).}
    \label{fig:spinup_amp3e5}
\end{figure}

For $\omega=0.100$ (Fig.~\ref{fig:urmax_amp3e5}, black curve), the system develops standing waves whose amplitude increases slowly. This is because the gradual spin-up of the core as a result of wave damping shifts the system closer to resonance with the mode that has an eigenfrequency of $\omega_{\text{eig}1}=0.09471$ in the absence of rotation. The 2D $u_\phi$ plots (Fig.~\ref{fig:2d_amp3e5}) for $t=4000$ and $t=8000$ show a standing wave on a background with a mean flow in the central region. That is why the appearance is modified compared with Fig. \ref{fig:2d_amp1e5}. The $u_r$ plot at $t=8000$ shows a slight reduction in the radial wavelength when compared with the $t=4000$ counterpart. Although difficult to see, this is a subtle but important effect, and we analyse more carefully later in paper.
The angular-velocity profile $\overline{\Omega}(r)$ gradually builds up (Fig.~\ref{fig:spinup_amp3e5}, left panel) and reaches the value of the pattern speed $\Omega_\text{p}=0.05$ at around $t \sim 10000$ (red line in Fig.~\ref{fig:spinup_amp3e5}). This leads to the formation of a critical layer and the wave is excluded from the central region. Thus, after reaching a peak value, $u_{r,\text{max}}$ drops dramatically as the motion is dominantly azimuthal, and waves exhibit largest $u_r$ further from centre, where the geometrical focusing is smaller.

For $\omega=0.102$ (Fig.~\ref{fig:urmax_amp3e5}, pink curve), the system also evolves into resonance with the same mode. Thus $u_{r,\text{max}}$ also initially increases, but at a slower rate than in the case $\omega=0.100$ because the resonance is more distant. The simulation eventually enters the critical-layer phase at around $t \sim 28000$.

As shown in Fig.~\ref{fig:urmax_amp3e5}, simulations with forcing frequencies $\omega=0.112$, $0.114$ and $0.116$ (blue, purple and brown curves, respectively) show similar behaviour: the wave amplitude increases as the system evolves closer to resonance with the mode that has an eigenfrequency of $\omega_{\text{eig}2}=0.11136$ in the absence of rotation. This behaviour occurs because these forcing frequencies are on the right-hand side of the closest eigenfrequency $\omega_{\text{eig}2}$. 

The forcing frequencies $\omega=0.110$, $0.108$, $0.106$ and $0.104$ (green, yellow, orange and red curves, respectively) are on the left-hand side of the closest eigenfrequency $\omega_{\text{eig}2}=0.11136$, 
and so in these cases the system evolves further away from resonance as a result of wave damping and spin-up. The wave amplitudes are gradually decreasing, as can be seen most clearly for $\omega=0.108$, $0.106$, and $0.110$ (yellow, orange and green curves, respectively). 

In Fig.~\ref{fig:peaks}, we show the different behaviours of the system starting with different forcing frequencies $\omega$. Red and blue arrows indicate the direction of evolution due to the spin-up of the core. The aforementioned behaviour of different $\omega$ can thus be understood (We remind the reader that the forcing frequency actually remains fixed in an inertial frame; in fact, it is the eigenfrequencies in an inertial frame that increase in time as a result of spin-up).

\subsubsection{High-amplitude forcing ($U=10^{-4}$, for which we expect $|A|=0.37$ at $\omega=0.118$)}

\begin{figure}
	\includegraphics[width=\columnwidth]{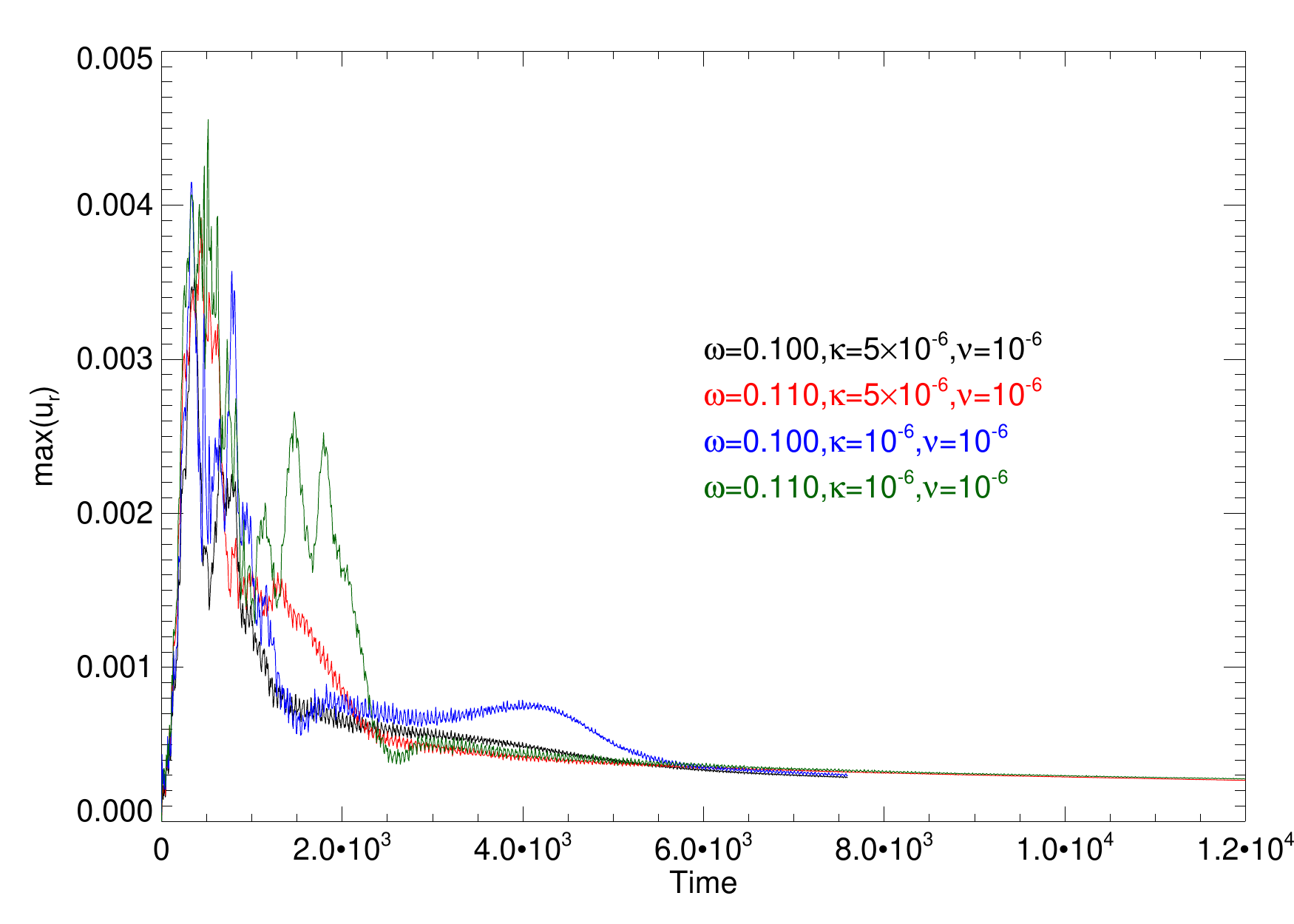}
    \caption{ $u_{r,\text{max}}$ versus time for simulations with high-amplitude forcing at frequencies $\omega=0.100$ and $\omega=0.110$. Two different values of $\text{Pr}$ 
    are considered.}
    \label{fig:urmax_amp1e4}
\end{figure}

In Fig.~ \ref{fig:urmax_amp1e4}, we show $u_{r,\text{max}}$ for four simulations with high-amplitude forcing. The general behaviour is an impulsive fast increase in the wave response which surpasses the threshold value for breaking very quickly within a few wave crossing times. The wave breaking leads to a dramatic drop in $u_{r,\text{max}}$ since the kinetic energy in the central region is 
dissipated and a mean azimuthal flow and critical layer (CL) is generated.

\begin{figure}
	\includegraphics[width=\columnwidth]{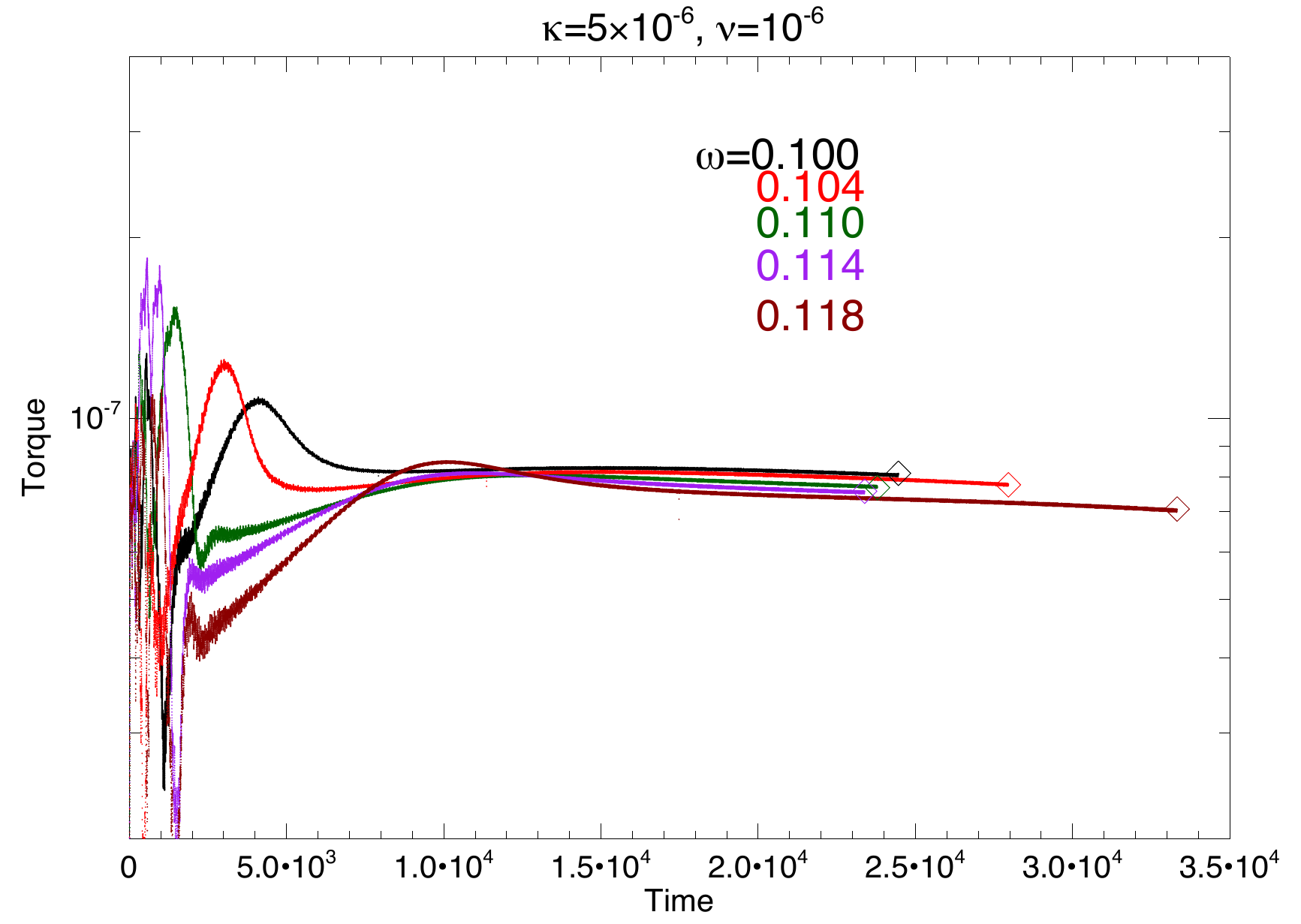}
    \caption{ Specific torque versus time for simulations with high-amplitude forcing
    and several forcing frequencies $\omega$.} 
    \label{fig:torque_amp1e4}
\end{figure}

\begin{figure*}
    \centering    \includegraphics[width=1.3\columnwidth,angle=90]{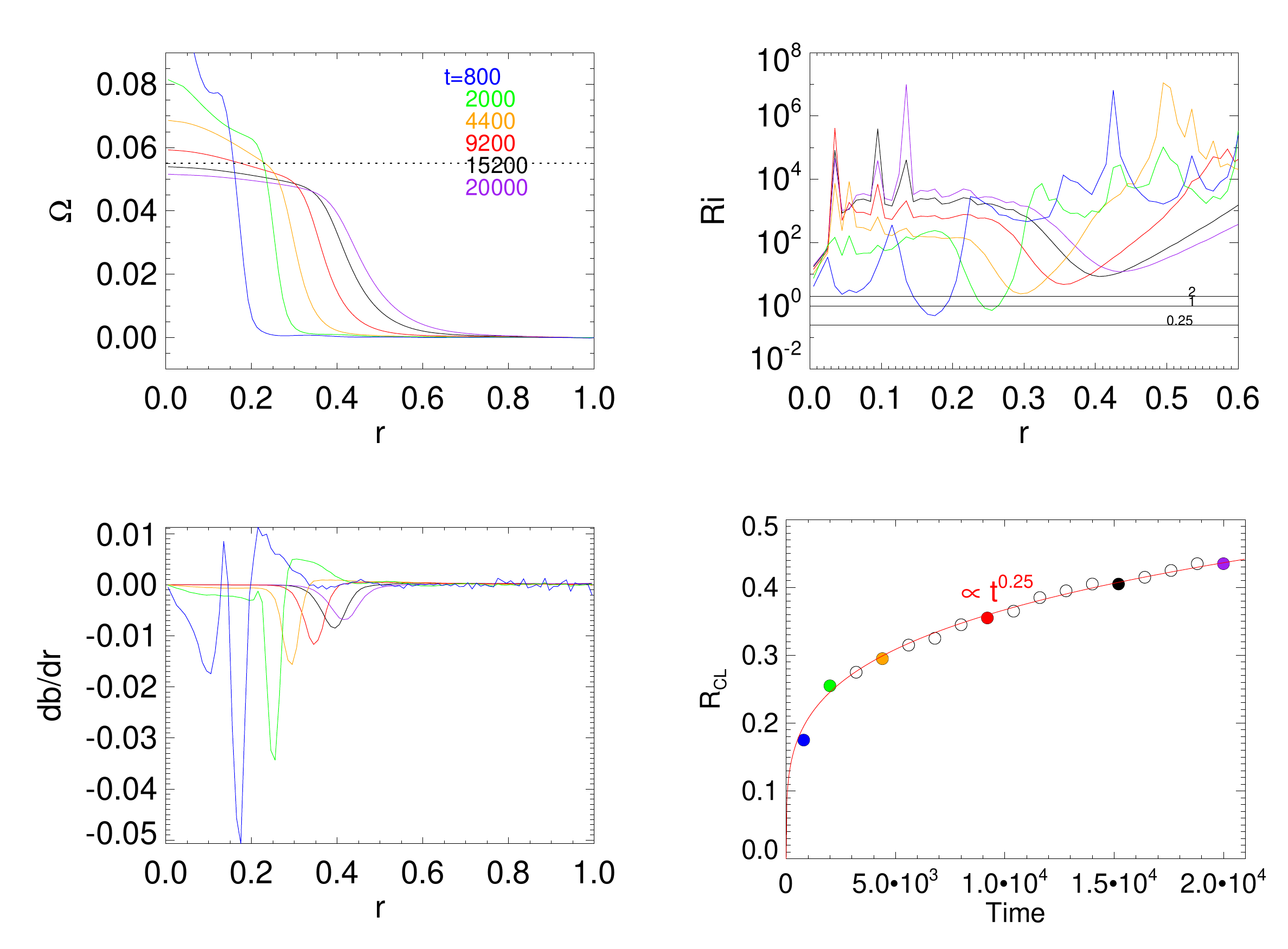}
    \caption{\textit{Upper left}: Azimuthally averaged angular-velocity profile $\overline{\Omega}(r)$, with the angular pattern speed shown as a dotted line; \textit{Lower left}: spatial derivative of the azimuthally averaged buoyancy $\dd\overline{b}/\dd r$; \textit{Upper right}: Richardson-number (Ri) profile; 
    \textit{Lower right}: radial position $R_\text{CL}$ of the critical layer versus time for different times: $t=800$ (blue), $2000$ (green)   $4400$ (orange), $9200$ (red), $15200$ (black)  and $20000$ (purple). This high-amplitude simulation  has the following parameters: $\omega=0.110$, $U=1\times10^{-4}$, $\nu=10^{-6}$ and $\kappa=5\times10^{-6}$. }
    \label{fig:Ri}
\end{figure*}

\begin{figure*}
    \centering    \includegraphics[width=2.0\columnwidth,angle=0]{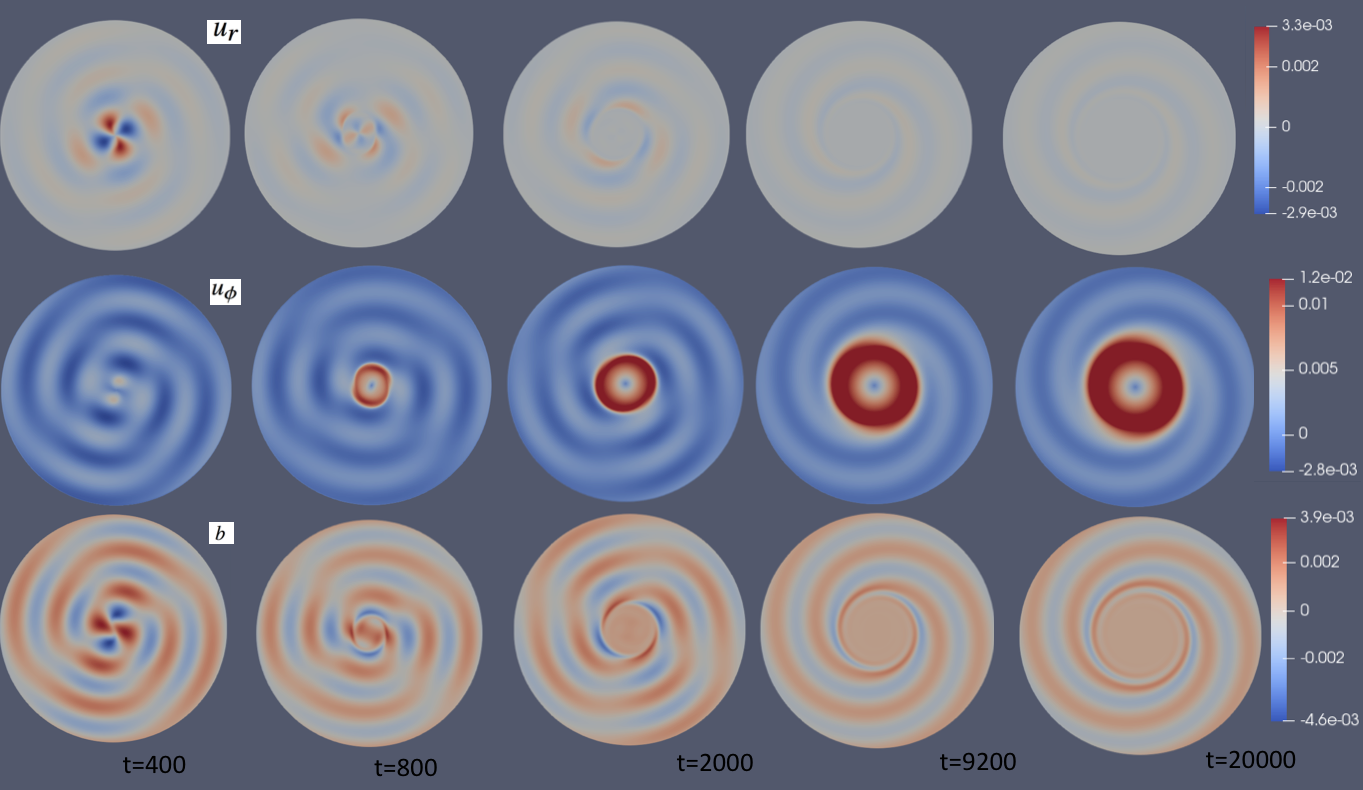}
    \caption{$u_r$, $u_{\phi}$ (radial and azimuthal velocities) and $b$ (buoyancy) for a simulation with high-amplitude forcing ($U=10^{-4}$) with frequency $\omega=0.11$. The diffusion coefficients are $\nu=10^{-6}$ and $\kappa=5\times 10^{-6}$. }

    
    \label{fig:2d_amp1e4}
\end{figure*}

As time progresses, the central region begins to spin up and the CL gradually moves outwards. 
We show the simulation results for a particular high-amplitude case 
($\omega=0.11$)
in Figs \ref{fig:Ri} and \ref{fig:2d_amp1e4}. In Fig.~\ref{fig:Ri}, the upper left panel shows the angular-velocity profile $\overline{\Omega}(r)$, which settles 
close to $\Omega_\text{p}=0.055$ inside a radius $R_\text{CL}(t)$ that advances outwards.

After wave breaking commences, the ingoing gravity waves deposit essentially all their AM flux at the CL (see the last three snapshots in Fig.~\ref{fig:2d_amp1e4} with $t \ge 2000$). 
If we assume that
the (expanding) spinning core  always rotates with the pattern speed $\Omega=\Omega_\text{p}$, and that the inward angular momentum flux $F$ can be treated as a constant, then 
we find a relation of the form $(\dd I/\dd t)\Omega_\text{p}=F$. Since the moment of inertia of a uniform disc of radius $R$ is $I=MR^2/2$ and its mass is $M=\pi R^2\rho$, where $\rho$ is the mass per unit area, we have $I\propto R^4$. We therefore expect the radial position of the CL to increase as $R_\text{CL}\propto t^{1/4}$.  A power-law fit to the graph of $R_\text{CL}$ vs $t$ is shown in the lower-right panel of Fig.~\ref{fig:Ri}. The resulting best-fitting power-law index of $0.2497$ is in excellent agreement with the above calculation. 
In Fig.~\ref{fig:torque_amp1e4}, we show the `tidal' specific torque $T_\text{s}$ in the simulations. This can be compared with the expected torque in the travelling-wave regime (equation~\ref{ttw}) after dividing by the mass of the fluid $\pi R^2 \rho$, i.e., $T_s \approx \frac{kR}{2}|U|^2 =\frac{1}{\omega} |U|^2$. With the high-amplitude forcing $|U|=10^{-4}$, and $\omega \approx 0.11$ as a typical value, we find $T_\text{s} \approx 9 \times 10^{-8}$, which is in good agreement with the values in the final slowly-varying stage in Fig.~\ref{fig:torque_amp1e4}. From the frequency-dependence of the values of the torque in these final stages, we find the torque does indeed roughly scale with $\omega^{-1}$ as expected (Fig.~\ref{fig:torque_peak}).

The Richardson number is a useful measure of the relative importance of stable stratification and shear. At the edge of the spinning core (defined as the location of a minimum in $\dd\overline{\Omega}/\dd r$) where the CL is found, we calculate the Richardson number as
$\text{Ri}=N^2/S^2$, where the shear rate is $S=r\,\dd\overline{\Omega}/\dd r$. The Brunt-V\"{a}is\"{a}l\"{a} frequency squared is $N^2=C^2r^2+r\,\dd\overline{b}/\dd r$. Note that the second term $r\,\dd\overline{b}/\dd r$ makes only a very small contribution to $N^2$ since $|\dd\overline{b}/\dd r| \ll r$ (lower-left panel of Fig.~\ref{fig:Ri}; see equation~\ref{background}). 

The calculated $\text{Ri}$ is shown in the upper right panel of Fig.~\ref{fig:Ri}. The location of the CL corresponds to a local minimum of $1/S^2$. It is very close to the minimum of $\text{Ri}$. The minimum $\text{Ri}$ ($\sim 0.5$) occurs for the blue curve. At later stages, $\text{Ri}$ in the CL is slightly larger than unity.

\citet{Su20} measured $\text{Ri}$ in their simulations
of upwardly propagating gravity waves in a plane-parallel atmosphere, and found that it
was close to the critical value of $1/4$ for the onset of a shear 
instability. Since we find that the minimum value of $\text{Ri}$ remains clearly above $1/4$ (the minimum $\text{Ri}$ is 0.49 as shown in the blue curve at $r \sim 0.17$), we do not believe that such an 
instability is limiting the sharpness of the CL in our simulations. 

\subsection{Effects of varying the Prandtl number}


\begin{figure}
	\includegraphics[width=\columnwidth]{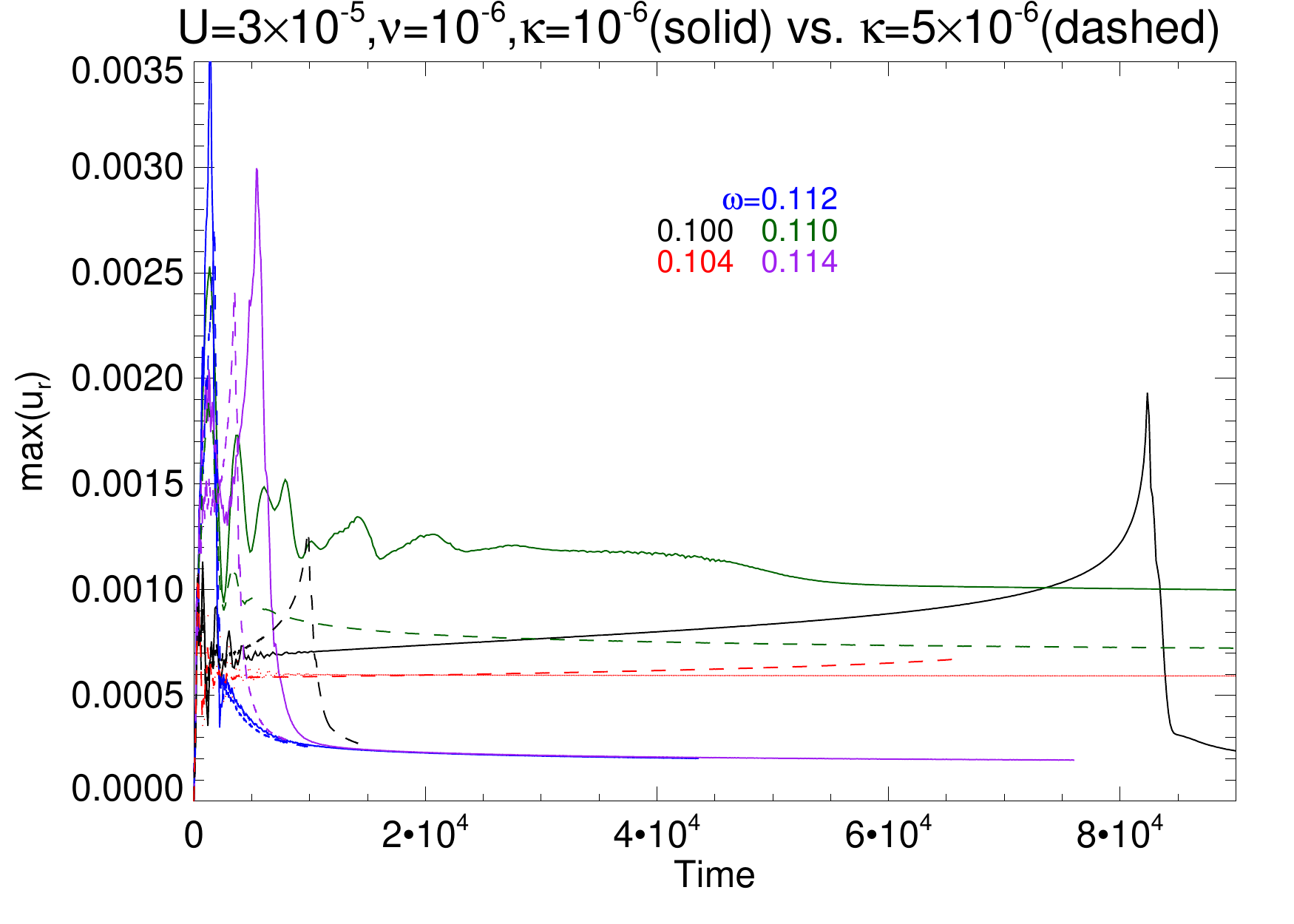}
    \caption{$u_{r,\text{max}}$ versus time for simulations with intermediate-amplitude forcing and two different Prandtl numbers. }
    \label{fig:amp3e5_bothPr}
\end{figure}

Bearing in mind that the Prandtl number $\text{Pr}=\nu/\kappa$ in the centre of a solar-type star is extremely small and beyond the reach of direct numerical simulations, we make a limited
exploration of the effects of varying $\text{Pr}$ by fixing $\nu=10^{-6}$ and varying $\kappa$. Thus $\kappa=10^{-6}$, $\kappa=5\times10^{-6}$ and $\kappa=1\times10^{-5}$ correspond to $\text{Pr}=1$, $\text{Pr}=0.2$ and $\text{Pr}=0.1$, respectively.

For the case of low-amplitude forcing, in which the system can evolve to an equilibrium angular velocity $\overline{\Omega}(r)$, since the spin-up rate $\partial\overline{\Omega}/\partial t$ due to wave damping 
is proportional to the total diffusivity $\nu+\kappa$, whereas the spin-down rate due to viscous diffusion is proportional to $\nu$, we expect larger $\overline{\Omega}$ for larger $\kappa$ given that $\nu$ is the same.
Fig.~\ref{fig:steady_omega} shows the dependence of equilibrium angular-velocity profile on the Prandtl number, 
based on equation~(\ref{omega_steady}).
The increasing trend of equilibrium angular velocity as $\text{Pr}$ is decreased is obvious.

\begin{figure}
	\includegraphics[width=\columnwidth]{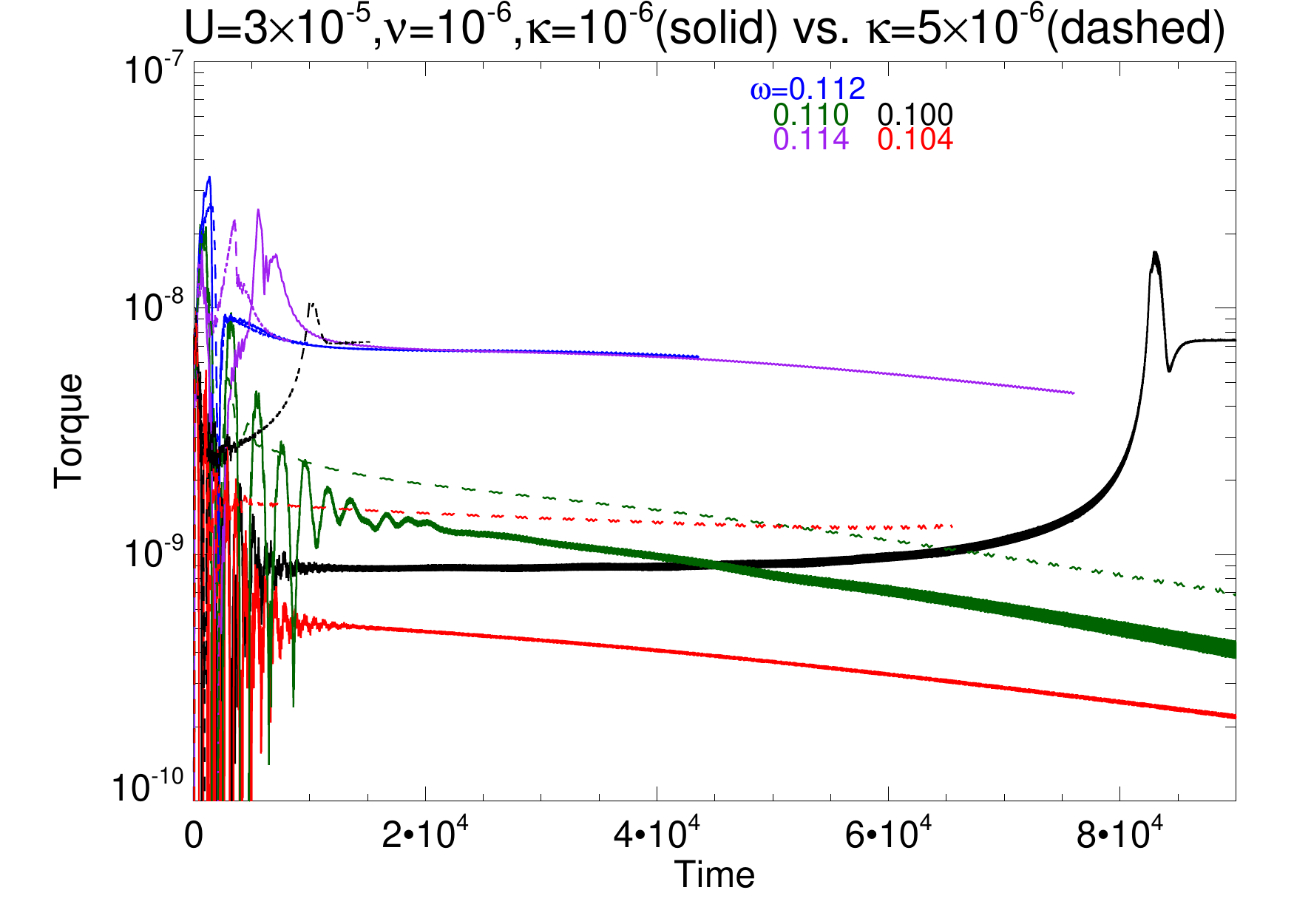}
    \caption{Specific torque versus time for simulations with intermediate-amplitude forcing and two different Prandtl numbers.}
    \label{fig:amp3e5_torque_bothPr}
\end{figure}

\begin{figure*}
   \centering
\includegraphics[width=1.6\columnwidth]{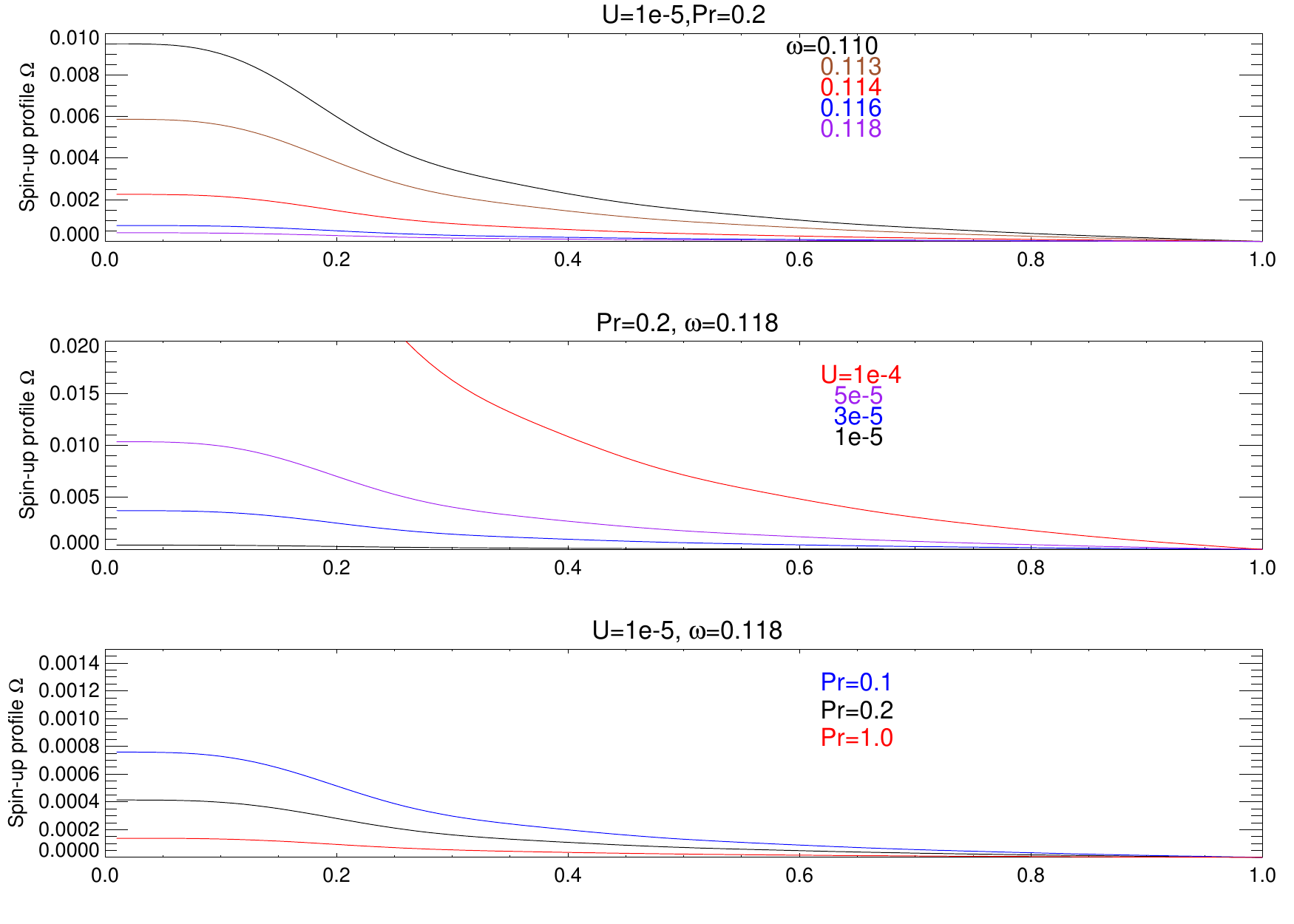}
    \caption{Equilibrium angular velocity profile $\overline{\Omega}(r)$ (based on equation~\ref{omega_steady}) for different forcing frequencies $\omega$, forcing amplitudes $U$ and Prandtl numbers $\text{Pr}$. These equilibrium solutions are self-consistent if $\overline{\Omega}\ll\Omega_\text{p}=\omega/2$.} 
    \label{fig:steady_omega}
\end{figure*}

For the case of intermediate-amplitude forcing, we compare the evolution of $u_{r,\text{max}}$ and the specific torque for $\text{Pr}=1$ and $\text{Pr}=0.2$ in Figs \ref{fig:amp3e5_bothPr} and \ref{fig:amp3e5_torque_bothPr}, respectively.
Simulations with smaller $\kappa$ (larger $\text{Pr}$) undergo longer transient phases because the total dissipation is weaker.
This can be seen, e.g., in the graphs for $\omega=0.104$ and $0.110$ in  Fig.~\ref{fig:amp3e5_bothPr}.

In Fig.~\ref{fig:amp3e5_torque_bothPr}, the green and red dashed lines ($\text{Pr}=0.2$) are systematically higher than their solid-line counterparts ($\text{Pr}=1$). Thus the torques are larger for the simulations with larger $\kappa$. Away from resonances and in the absence of wave breaking or critical layers, the deposition of AM scales with the total diffusivity $\nu+\kappa$.

The larger $\text{Pr}$ simulations, with lower $\kappa$, take much longer times to spin up the core to the pattern speed $\Omega_\text{p}$. For forcing frequency $\omega=0.100$ in Fig.~\ref{fig:amp3e5_torque_bothPr}, the $\text{Pr}=1$ ($\kappa=10^{-6}$) case (black solid line) takes about $t=8\times10^{4}$ to spin up the core to $\Omega=0.05$, while the $\text{Pr}=0.2$ ($\kappa=5\times10^{-6}$) case (dashed solid line) only needs about $t=1\times10^{4}$. In Section~\ref{s:implications}, we discuss the implications of this trend in the application to real stars. 
For forcing frequencies $\omega=0.114$ and $0.112$, which are close to an eigenfrequency, wave breaking/critical-layer formation occurs promptly.

\section{Analytical approach}
\label{s:analytical}

\subsection{Streamfunction and vorticity}

To develop our analytical approach to the problem, we first rewrite the basic equations (\ref{du})--(\ref{divu}) in a scalar form. Following a standard procedure for 2D Boussinesq or incompressible flows, we introduce the streamfunction $\psi$ and vorticity $\zeta$ such that $\bm{u}=\bm{\nabla}\bm{\times}(\psi\,\bm{e}_z)$ and $\bm{\nabla}\bm{\times}\bm{u}=\zeta\,\bm{e}_z=-\nabla^2\psi\,\bm{e}_z$, where $\bm{e}_z$ is a unit vector perpendicular to the plane of the flow. Equation~(\ref{divu}) is then automatically satisfied, while equations (\ref{du}) and (\ref{db}) can be reformulated (after taking the curl of the former to eliminate the pressure) as
\begin{align}
  &\frac{\partial\zeta}{\partial t}+\frac{\partial b}{\partial\phi}=\nu\nabla^2\zeta+J(\psi,\zeta),\label{dzeta}\\
  &\frac{\partial b}{\partial t}+C^2\frac{\partial\psi}{\partial\phi}=\kappa\nabla^2b+J(\psi,b),\\
  &\zeta=-\nabla^2\psi,\label{zeta}
\end{align}
in which the nonlinear terms involve the Jacobian operator $J$. In polar coordinates, and for any functions $f(r,\phi,t)$ and $g(r,\phi,t)$,
\begin{align}
  &u_r=\frac{1}{r}\frac{\partial\psi}{\partial\phi},\qquad
  u_\phi=-\frac{\partial\psi}{\partial r},\\
  &\nabla^2f=\frac{1}{r}\frac{\partial}{\partial r}\left(r\frac{\partial f}{\partial r}\right)+\frac{1}{r^2}\frac{\partial^2f}{\partial\phi^2},\\
  &J(f,g)=\frac{1}{r}\left(\frac{\partial f}{\partial r}\frac{\partial g}{\partial\phi}-\frac{\partial g}{\partial r}\frac{\partial f}{\partial\phi}\right).
\end{align}

\subsection{Linear waves on the original background}
\label{s:linear_original}

The original basic state of our model is non-rotating and has a stable stratification with buoyancy frequency $N=Cr$. This corresponds to the trivial solution $\psi=\zeta=b=0$ of equations (\ref{dzeta})--(\ref{zeta}). 
Linear waves on this background satisfy the same equations without the nonlinear Jacobian terms. We find wave solutions of the form
\begin{align}
  &\psi=\text{Re}\left[\alpha f(r)\exp(\ii m\phi-\ii\omega t)\right],\\
  &b=\text{Re}\left[\beta f(r)\exp(\ii m\phi-\ii\omega t)\right],
\end{align}
where $m$ is the (integer) azimuthal wavenumber, $\omega$ is the (real) angular frequency, $\alpha$ and $\beta$ are (complex) amplitudes and $f(r)$ is an eigenfunction of the transformed Laplacian operator such that
\begin{equation}
  \mathcal{L}f:=-\frac{1}{r}\frac{\dd}{\dd r}\left(r\frac{\dd f}{\dd r}\right)+\frac{m^2f}{r^2}=k^2f,\label{bessel}
\end{equation}
where $k$ is a constant. Equations (\ref{dzeta})--(\ref{zeta}) are satisfied if
\begin{align}
  &(-\ii\omega+\nu k^2)k^2\alpha+\ii m\beta=0,\\
  &(-\ii\omega+\kappa k^2)\beta+\ii mC^2\alpha=0,
\end{align}
leading to the dispersion relation \citep{Bar11b}
\begin{equation}
  (\omega+\ii\nu k^2)(\omega+\ii\kappa k^2)k^2=m^2C^2.\label{dr}
\end{equation}
The solution of equation~(\ref{bessel}) that is regular at $r=0$ is the Bessel function of the first kind, $f\propto J_m(kr)$.

In the absence of diffusion, equation~(\ref{dr}) implies $k=C/\Omega_\text{p}$, as stated in Section~\ref{s:gravity}. For forced waves with real frequency $\omega$ in the presence of diffusion, equation~(\ref{dr}) is a cubic for $k^2$ and has three complex roots. Either square root of $k^2$ can be taken, because $J_m(-kr)=(-1)^mJ_m(kr)$. If diffusion is weak ($\nu,\kappa\ll\omega^3/m^2C^2$), as in the case of stellar applications, then one of the roots for $k^2$ has a small imaginary part and represents a standing wave slightly affected by diffusion. The other two roots have large imaginary parts and represent strongly attenuated disturbances that are important only in a thin layer near the outer boundary.

Strictly speaking, the forced wave problem with diffusion requires three outer boundary conditions, such as the specification of $u_r$, $u_\phi$ and $b$. In practice, when diffusion is weak, the boundary condition on $u_r$ (or equivalently $\psi$) determines the amplitude of the standing wave, as in the ideal case, while the remaining two boundary conditions determine the amplitudes of the boundary-layer corrections described above.

\subsection{Linear waves on a slowly evolving background}
\label{s:linear}

In the numerical simulations in which the internal gravity waves do not break promptly, the waves propagate on a background state that evolves gradually: it develops a mean flow (also known as a zonal flow, or differential rotation) and the stratification changes.
Interactions of waves and mean flows are familiar, especially in geophysical fluid dynamics \citep{Buh14}. 

We therefore generalize the linear analysis of the previous subsection by considering an evolved basic state with streamfunction $\overline{\psi}(r)$ and buoyancy $\overline{b}(r)$. The corresponding angular velocity, vorticity and buoyancy frequency are given by
\begin{equation}
  \overline{\Omega}=-\frac{1}{r}\frac{\dd\overline{\psi}}{\dd r},\qquad
  \overline{\zeta}=\frac{1}{r}\frac{\dd}{\dd r}\left(r^2\overline{\Omega}\right),\qquad
  N^2=C^2r^2+r\frac{\dd\overline{b}}{\dd r}.\label{background}
\end{equation}
We then consider linear perturbations on this background, neglecting its slow evolution in time, such that
\begin{align}
  &\psi=\overline{\psi}(r)+\text{Re}\left[\psi'(r)\exp(\ii m\phi-\ii\omega t)\right],\label{psi_decomp}\\
  &b=\overline{b}(r)+\text{Re}\left[b'(r)\exp(\ii m\phi-\ii\omega t)\right].\label{b_decomp}
\end{align}
With this assumption, the quantities denoted by an overbar correspond to the azimuthally averaged variables considered in Section~\ref{s:simulations}.

The wave equations governing the perturbations are obtained by linearizing equations (\ref{dzeta})--(\ref{zeta}):
\begin{align}
  &\ii m\left(\overline{\Omega}-\Omega_\text{p}\right)\zeta'+\ii m\frac{1}{r}\frac{\dd\overline{\zeta}}{\dd r}\psi'+\ii mb'=-\nu\mathcal{L}\zeta',\label{linear1}\\
  &\ii m\left(\overline{\Omega}-\Omega_\text{p}\right)b'+\ii m\frac{N^2}{r^2}\psi'=-\kappa\mathcal{L}b',\\
  &\zeta'=\mathcal{L}\psi'.\label{linear3}
\end{align}
We discuss the nonlinear feedback of the waves and the resulting evolution of the background state in Section~\ref{s:evolution} below. 

The linearised equations for waves can also be expressed in terms of the primitive variables: the velocity perturbations $u'_r$ and $u'_{\phi}$, pressure perturbation $q'$ and buoyancy perturbation $b'$: 

\begin{align}
&\ii m\left(\overline{\Omega}-\Omega_\text{p}\right) u'_r - 2\overline{\Omega} u'_{\phi} = -\frac{\dd q'}{\dd r} +rb'\nonumber\\
&\qquad+ \nu \left [\frac{1}{r} \frac{\dd}{\dd r}\left(r\frac{\dd u'_r}{\dd r}\right) -\left(m^2+1\right)\frac{u'_r}{r^2} -\frac{2\ii m}{r^2}u'_{\phi} \right],\label{primitive1}
\end{align}
\begin{align}
&\ii m\left(\overline{\Omega}-\Omega_\text{p}\right) u'_{\phi}+\frac{u'_r}{r}\frac{\dd}{\dd r}\left(r^2\overline{\Omega}\right)=-\frac{\ii m}{r}q'\nonumber\\
&\qquad+ \nu \left[\frac{1}{r}\frac{\dd}{\dd r}\left(r\frac{\dd u'_{\phi}}{\dd r}\right)-\left(m^2+1\right)\frac{u'_{\phi}}{r^2}+\frac{2\ii m}{r^2} u'_r \right]
\end{align}
\begin{align}
&\ii m\left(\overline{\Omega}-\Omega_\text{p}\right) b' + u'_r \left (C^2 r +\frac{\dd\overline{b}}{\dd r} \right) =\kappa \left [\frac{1}{r} \frac{\dd}{\dd r}\left(r\frac{\dd b'}{\dd r}\right)-m^2\frac{b'}{r^2} \right],\\
&\frac{1}{r}\frac{\dd}{\dd r}(ru'_r) + \frac{\ii m}{r}u'_{\phi} =0.\label{primitive4}
\end{align}

We solve these ordinary differential equations numerically for background states $\overline{\Omega}(r)$ and $\overline{b}(r)$ extracted from the simulations. A Chebyshev spectral collocation method \citep{boy01} similar to that of \citet{Ogi04} is adopted with 200 Chebyshev-Gauss-Lobatto points. 
We use stress-free boundary conditions at a small radius $r_\text{in}=0.001$, i.e., $u'_r=0$,
$(\dd/\dd r)(u'_{\phi}/r)=0$
and $b'=0$, and the same outer boundary conditions as in the simulations (equation~\ref{bcs_simulation}): $u_r'=U$, $u'_\phi=0$ and $b'=-\ii U/\omega$ at $r=1$ (in units such that $R=C=1$).
Note that for the pressure perturbation $q'$, we replace the corresponding boundary condition with the incompressibility condition~(\ref{primitive4}).

\begin{figure}
	\includegraphics[width=\columnwidth]{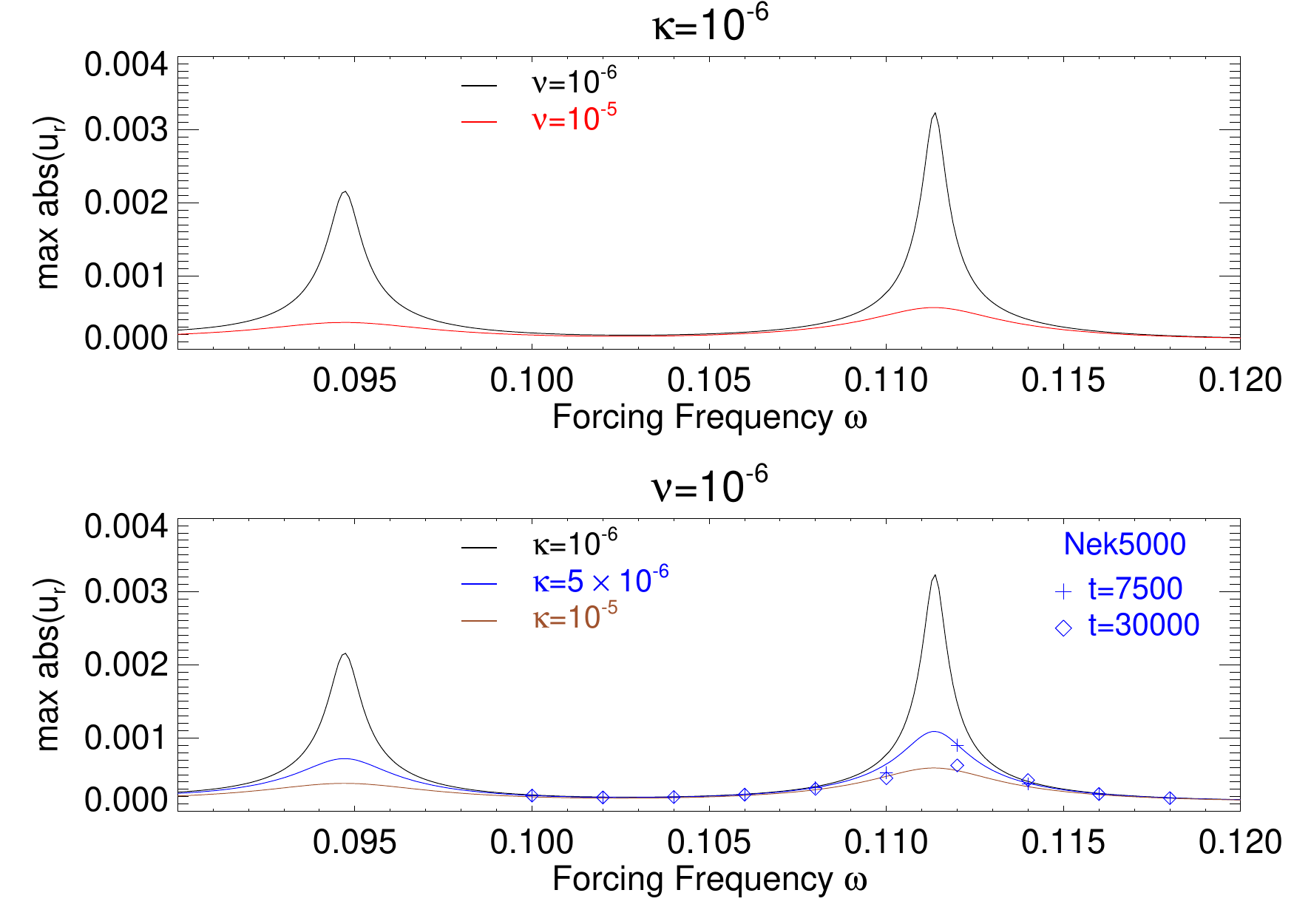}
    \caption{Responses of linear gravity waves on the original background. The maximum radial velocity $u_{r,\text{max}}$ occurring in solutions of the linearized equations (\ref{primitive1})--(\ref{primitive4}) with low-amplitude forcing is shown as a function of the forcing frequency $\omega$ for two different viscosities $\nu=10^{-6}$ and $10^{-5}$ (upper panel, black and red curves) and for different thermal diffusivities $\kappa=10^{-6}$, $5\times 10^{-6}$, and $10^{-5}$ (lower panel, black, blue and brown curves). In each case the two peaks correspond to resonances with standing modes having eigenfrequencies of $\omega_{\text{eig}1}=0.09471$ and $\omega_{\text{eig}2}=0.11136$. The corresponding values of $u_{r,\text{max}}$ from Nek5000 simulations with low-amplitude forcing are indicated by blue symbols. Crosses and diamonds indicate values at $t=7500$ and $t=30000$, respectively.}
    \label{fig:reso_peak}
\end{figure}

We first solve the linearized equations on the original, non-rotating background. The solutions obtained in this case are equivalent to the analytical solutions involving (complex) Bessel functions described in Section~\ref{s:linear_original}. The `tidal' response as a function of the forcing frequency is shown in Fig.~\ref{fig:reso_peak} for different values of the diffusivities. The response curves show  Lorentzian-shaped resonance peaks, each resembling the response of a lightly damped harmonic oscillator. The figure also shows the expected behaviour that away from (close to) resonance there is a weak (strong) dependence on the diffusivity.
Note that appropriate outer boundary conditions are needed to obtain this behaviour. Increasing the viscosity or thermal diffusivity results in broader and lower resonance peaks. In the lower panel, the responses measured in the low-amplitude Nek5000 simulations are indicated by the blue crosses and diamonds, corresponding to early ($t=7500$) and later ($t=30000$) stages, respectively. Note that away from resonance, the crosses match the linear wave response (blue curve) well. However, for forcing frequencies close to resonances, owing to a greater spinning up of the core, the Nek5000 values are different from the linear wave expectations with the original background. This can be seen more obviously for $\omega=0.110$ and $0.112$, for which from $t=7500$ to $t=30000$ the system evolves away from resonance and the responses evolve to lower values.

\begin{figure*}
\centering
	\includegraphics[width=1.4\columnwidth,angle=0]{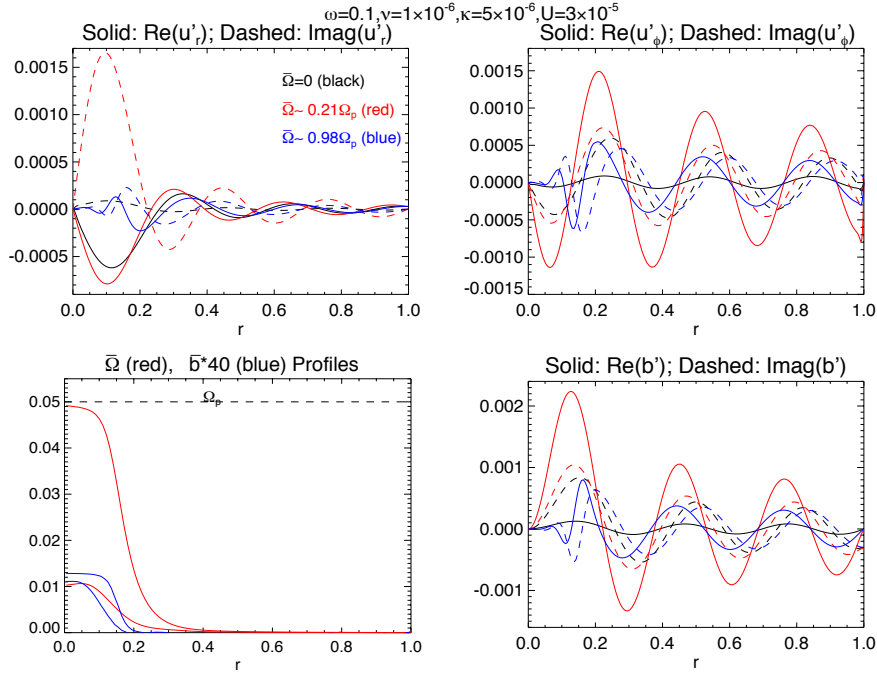}
    \caption{Linear wave solutions for $u'_r$, $u'_{\phi}$ and $b'$ with three different background states taken from a nonlinear simulation at $t_1=0$, $t_2=8350$ and $t_3=15260$. The corresponding central angular velocities are $\overline{\Omega}(0)=0$, $0.21\,\Omega_\text{ p}$ and $0.98\,\Omega_\text{p}$. The simulation has parameters $U=3\times 10^{-5}$, $\omega=0.100$, $\nu=10^{-6}$ and $\kappa=5\times 10^{-6}$. The lower left panel shows the azimuthally averaged profiles of angular velocity $\overline{\Omega}$ (red) and buoyancy $\overline{b}$ (blue) at the two later times $t_2$ (lower) and $t_3$ (higher), with the angular pattern speed indicated by the horizontal dashed line.   }
    \label{fig:linsol_2bkg}
\end{figure*}

We then examine how the rotating background modifies the linear waves. The linear wave solutions $u'_r$, $u'_{\phi}$ and $b'$ are shown in Fig.~\ref{fig:linsol_2bkg} for two different background states with central angular velocities of $\overline{\Omega}(0)=0.21\,\Omega_\text{p}$
(red) and $0.98\,\Omega_\text{p}$ 
(blue). The profiles of  angular velocity and buoyancy are extracted from a simulation with intermediate-amplitude forcing at $t=8350$ and $t=15260$, respectively. The corresponding solution without background rotation ($\overline{\Omega}=0$, $t=0$) is also shown as black curves.

From the upper-left panel, we can see that the wavelength of the $u'_r$ profile becomes shorter for the $\overline{\Omega}(0)=0.21\,\Omega_\text{p}$ profile (red) when compared with the non-rotating background solution (black). This is because the wave frequency becomes smaller in the frame rotating with the fluid, and the dispersion relation of gravity modes implies a shorter wavelength. Similar wavelength-shortening can also be seen in the $u'_{\phi}$ and $b'$ profiles.

The second background state (blue)
corresponds to a very late stage in which the core has been already spun up to the pattern speed, meaning that a critical layer is formed near the centre. The solution is now in the travelling-wave regime.
This can be seen from the $\pi/2$ phase difference between the real part (solid blue) and imaginary part (dashed blue) of the $u'_r$, $u'_{\phi}$ and $b'$ profiles.
We can also see that an additional wavelength appears in the inner region ($r < 0.2$). In this case the shortening of the wavelength is sufficient for the wave to be absorbed in the critical layer as a result of viscosity and thermal diffusion.

In this framework for wave--mean-flow interactions, the waves are assumed to be linear, no wave-wave interactions are considered and wave breaking is not included. However, we find the wave profiles calculated from equations (\ref{primitive1})--(\ref{primitive4}) with an evolved background state in which the central angular velocity is close to $\Omega_\text{p}$ can match very well the waves in the fully nonlinear Nek5000 simulations.
For example, the traveling-wave profiles of $u'_r$ and $u'_{\phi}$ in Fig.~\ref{fig:linsol_2bkg} for $\overline{\Omega}(0)=0.98\,\Omega_\text{p}$ are essentially the same as the Nek5000 results (not shown). Thus this quasi-linear approach can still be used to model the spin-up of the core from zero to the pattern speed, as well as the absorption of the ingoing waves by the critical layer, even if the mechanism of absorption is not necessarily the same.

We find that the effect on the waves of the altered stratification in the evolved background states, described by the azimuthally averaged buoyancy perturbation $\overline{b}(r)$, is relatively unimportant. In fact, the buoyancy profile is not significantly altered even in the large-amplitude forcing case. This can be seen from Fig.~\ref{fig:linsol_2bkg} and the $N^2$ expression in Eq. \ref{background}.


\subsection{Evolution of the slowly evolving background}
\label{s:evolution}

If the perturbations are not too large, we can use the linear wave solutions to estimate the gradual evolution of the background state, i.e.\ the profiles of $\overline{\Omega}(r,t)$ and $\overline{b}(r,t)$. 
From the conservation laws for angular momentum and buoyancy (Section~\ref{s:conservation}) we deduce that the mean angular velocity and buoyancy evolve according to
\begin{align}
  &\frac{\partial\overline{\Omega}}{\partial t}=\frac{1}{r^3}\frac{\partial}{\partial r}\left(r^3\nu\frac{\partial\overline{\Omega}}{\partial r}\right)+S_\Omega,\label{domegadt}\\
  &\frac{\partial\overline{ b}}{\partial t}=\frac{1}{r}\frac{\partial}{\partial r}\left(r\kappa\frac{\partial\overline{b}}{\partial r}\right)+S_b,\label{dbdt}
\end{align}
with source terms
\begin{align}
  &S_\Omega=-\frac{1}{r^3}\frac{\partial}{\partial r}\left[\frac{1}{2}r^2\,\text{Re}\left(u_r'^*u_\phi'\right)\right],\\
  &S_b=-\frac{1}{r}\frac{\partial}{\partial r}\left[\frac{1}{2}r\,\text{Re}\left(u_r'^*b'\right)\right].
\end{align}

In terms of the streamfunction, vorticity and buoyancy perturbations, we find, by manipulating the linearized equations (\ref{linear1})--(\ref{linear3}), that
\begin{align}
  &S_\Omega=\frac{1}{2r^2}\left[\frac{\nu\,\text{Re}\left(\psi'^*\mathcal{L}\zeta'\right)}{\left(\Omega_\text{p}-\overline{\Omega}\right)}+\frac{\kappa\,\text{Re}\left(\psi'^*\mathcal{L}b'\right)}{\left(\Omega_\text{p}-\overline{\Omega}\right)^2}\right],\\
  &S_b=\frac{1}{r}\frac{\dd}{\dd r}\left[\frac{\kappa\,\text{Re}\left(\psi'^*\mathcal{L}b'\right)}{2\left(\Omega_\text{p}-\overline{\Omega}\right)}\right].
\end{align}
These expressions show explicitly that the source terms depend on diffusion. In the absence of diffusion, the phase relationships between $u_r'$, $u_\phi'$ and $b'$ are such that the fluxes of angular momentum and buoyancy vanish.

If the source terms are estimated using the ideal linear wave solution (for $m=2$) on the original background with $\overline{\Omega}=\overline{b}=0$ (Section~\ref{s:gravity}), then we have (in terms of the dimensionless radial variable $x=kr$)
\begin{align}
  &S_\Omega=8|A|^2\Omega_\text{p}(\nu+\kappa)k^2\,\frac{J_2(x)^2}{x^2},\label{s_omega_estimate}\\
  &S_b=8|A|^2\Omega_\text{p}^2\kappa k^2\,\frac{1}{x}\frac{\dd}{\dd x}\left[J_2(x)^2\right],\label{s_b_estimate}
\end{align}
with a linear dependence on the diffusion coefficients. To describe the radial profile of the spin-up process, we define the normalized dimensionless function
\begin{equation}
  f_1(x)=30.88\,\frac{J_2(x)^2}{x^2},\label{f1}
\end{equation}
which has a peak value of $1$ at $x=2.300$.

\begin{figure}
	\includegraphics[width=\columnwidth]{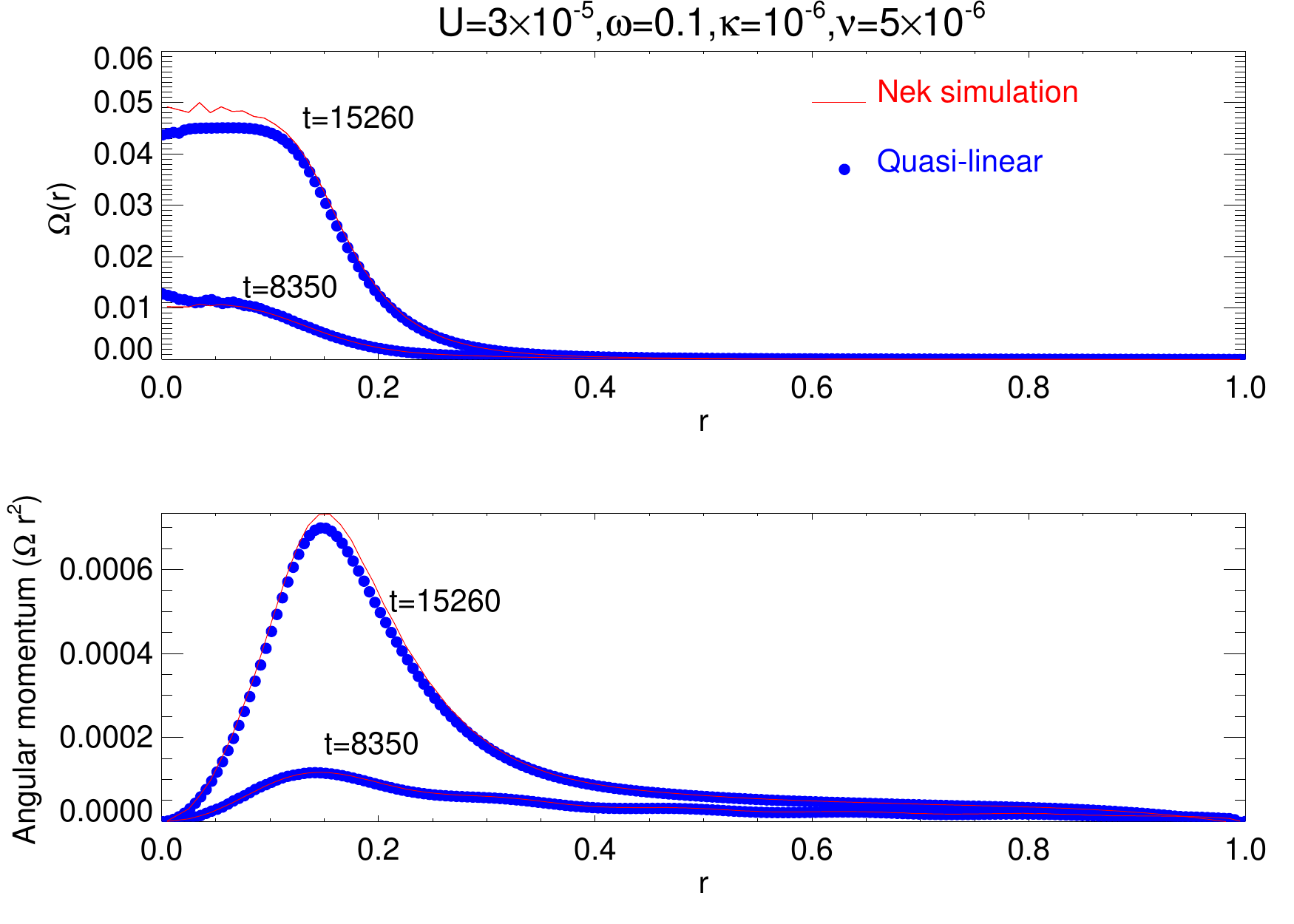}
    \caption{Angular velocity and angular momentum profiles at two different moments $t=8350$ and $t=15260$. Blue and red curves indicate the quasi-linear and full nonlinear calculations, respectively.}
    \label{fig:evolve_quasilinear}
\end{figure}

In this quasi-linear approximation \citep[e.g.,][]{bis93}, the waves are assumed to be linear and described by equations (\ref{primitive1})--(\ref{primitive4}). Meanwhile, the background is slowly evolving because of diffusion and the source terms resulting from wave damping (equations \ref{domegadt} and \ref{dbdt}). 
To describe the evolution of the background far from the original state, we 
iteratively solve the evolutionary equations (\ref{domegadt}) and (\ref{dbdt}) along with the wave equations (\ref{primitive1})--(\ref{primitive4}). The wave equations are solved with the Chebyshev collocation method and the background state is updated using the Crank--Nicolson time-stepping method. The source terms $S_{\Omega}$ and $S_b$ are updated every $\Delta t= 500$. Fig.~\ref{fig:evolve_quasilinear} shows the radial profiles of angular velocity $\overline{\Omega}$ and angular momentum $r^2\overline{\Omega}$ at two different moments $t=8350$ and $t=15260$. Compared with the fully nonlinear Nek5000 simulation, the quasi-linear method yields satisfactory results, even when the angular velocity is very close to the pattern speed. The small difference at $t=15260$ is likely due to the fact that the source term at this later stage would need more frequent updates than at the early stage. Note that even though nonlinear processes such as wave-wave interactions and wave breaking cannot be studied in the quasi-linear approximation, this method is still successful in producing the critical-layer formation.

\subsection{Equilibrium solutions involving small changes in the background}
\label{s:equilibrium}

If we estimate the source terms as in equations (\ref{s_omega_estimate})--(\ref{s_b_estimate}), then we can solve equations (\ref{domegadt})--(\ref{dbdt}) analytically in a steady state to find an equilibrium solution for the evolved background state. 

Applying a regularity condition at $r=0$, we find
\begin{align}
  &\frac{\overline{\Omega}}{\Omega_\text{p}}=\left(1+\frac{1}{\text{Pr}}\right)|A|^2f_2(x),\label{omega_steady}\\
  &\frac{r\frac{\dd\overline{b}}{\dd r}}{C^2r^2}=-8|A|^2\frac{J_2(x)^2}{x^2}=-0.2591|A|^2f_1(x),\label{b_steady}
\end{align}
where
\begin{equation}
  f_2=J_0^2+4J_1^2-4J_0J_2-J_2^2
\label{f2}
\end{equation}
is a positive dimensionless function composed of products of Bessel functions\footnote{We found it helpful to use the recurrence relation $(2/x)J_1(x)=J_0(x)+J_2(x)$.}, which satisfies $f_2(0)=1$ and decreases monotonically towards $0$ as $x\to\infty$. [This is the solution for an unbounded domain in which $\overline{\Omega}\to0$ as $r\to\infty$; if instead the outer boundary condition requires $\overline{\Omega}$ to vanish at $r=R$, then the relevant solution is obtained by subtracting a constant angular velocity, i.e.\ by replacing $f_2(x)$ with $f_2(x)-f_2(X)$.] 

The equilibrium solution involves a prograde differential rotation, representing a balance between the diffusive deposition of angular momentum by the waves and the outward viscous transport of angular momentum, together with a reduction of the stable stratification, resulting from the diffusive mixing of entropy.
The steady-state angular-velocity profile has an important dependence on the Prandtl number, while the buoyancy profile does not; this is because the buoyancy source term requires thermal diffusion, while the angular-momentum source term has contributions from both viscosity and thermal diffusion.

Equations (\ref{omega_steady}) and (\ref{b_steady}) are written in a way that expresses the fractional change in the background state. In order for this solution to be self-consistent, the changes in the background state should be small so that their effect on the wave can be neglected. Since the maximum value of $|J_2(x)/x|$ is $0.1800$ (and occurs at $x=2.300$), the fractional change in $N^2$ is everywhere less than $0.259|A|^2$, which in turn is significantly less than unity for waves below the breaking amplitude. However, the predicted $\overline{\Omega}/\Omega_\text{p}$ can exceed unity even if $|A|<1$, especially if the Prandtl number is very small, as it is in stars.

This analysis suggests that the most important effect of the dissipation of subcritical waves, especially at low $\text{Pr}$, is to modify the angular-velocity profile in the central region. For very low $\text{Pr}$, there is a broad range of subcritical wave amplitudes that may allow the formation of a critical layer. Of course, if $\overline{\Omega}/\Omega_\text{p}$ is not small then this analysis is not self-consistent and the solution above cannot be trusted; the effect of spin-up on the waves needs to be taken into account as we did numerically in Section~\ref{s:evolution}.

We examine how the equilibrium angular-velocity profile  varies with the forcing frequency, forcing amplitude and Prandtl number in Fig.~\ref{fig:steady_omega}.

To estimate the time-scale on which the equilibrium solution (where this is valid) is established,
we can write the angular-velocity profile as
the sum of the equilibrium profile and a time-dependent correction $\tilde\Omega(r,t)$, which then satisfies the homogeneous version of the diffusion equation (\ref{domegadt}) without any source term.
By separation of variables, this equation admits solutions of the form $\tilde{\Omega}=f(r)\exp{(-\lambda t)}$,
where $f(r) \propto J_1(Kr)/r$ and $\lambda=\nu K^2$. The outer boundary condition requires $\tilde\Omega(R,t)=0$, so $KR$ must be one of the zeros of $J_1(x)$, the smallest of which is $x_1=3.832$. (Note that $K$ here differs from the wavenumber $k$ of the gravity wave used elsewhere in the paper.) The largest-scale mode therefore decays exponentially on a time-scale $\lambda^{-1}=(1/x_1^2)R^2/\nu=0.06811\,R^2/\nu \sim l^2_c/\nu$, which is $68110$ in the case $R=1$, $\nu=10^{-6}$. This is comparable with a viscous timescale on a length-scale $l_c \approx 0.26R$, which is approximately the size of the differentially-rotating ``core" in the figures e.g. Fig. \ref{fig:steady_amp1e5}). We expect the equilibrium angular-velocity profile to be approached on this time-scale.

A similar treatment can be applied to the buoyancy, in which case the solution for $\tilde b$ involves $J_0$ rather than $J_1$. The slowest-decaying mode for the equilibration of the buoyancy profile has a time-scale $0.1729\,R^2/\kappa$, which is $34580$ in the case $R=1$, $\kappa=5\times 10^{-6}$. This can explain why equilibration has occurred for $\overline{b}$ in the examples shown in Fig. \ref{fig:steady_amp1e5}, for example, unlike for the mean flow $\overline{\Omega}$).

\subsection{Evolution of the specific torque}
\label{s:evolution_torque}

In Figs \ref{fig:torque_amp1e5}, \ref{fig:torque_amp3e5} and \ref{fig:torque_amp1e4} we show the evolution of the specific torque, 
$T_\text{s}=\frac{\dd L}{\dd t}=\int_0^R r^2(\partial\overline{\Omega}/\partial t)\, 2\pi r\,\dd r / \pi R^2$.

\begin{figure}
	\includegraphics[width=\columnwidth]{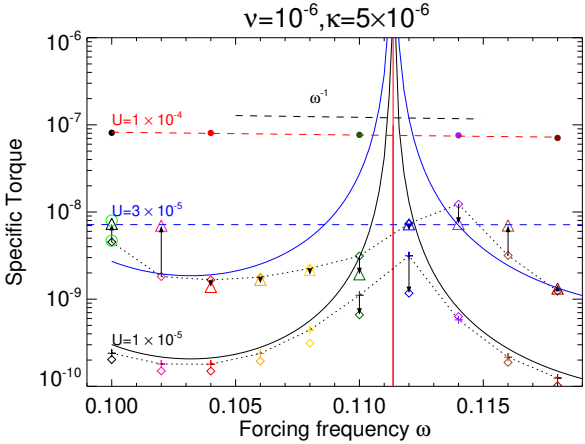}
    \caption{Specific torque for the high, intermediate, and low amplitude forcing cases. High-amplitude forcing ($U=10^{-4}$) cases always lead to traveling waves, indicated by filled circles. A power law $\propto \omega^{\alpha}$ fit (red dashed line) yields $\alpha=-0.85$, which is close to the expected theoretical dependence $\propto \omega^{-1}$ indicated by the short, black dashed line. Intermediate ($U=3\times 10^{-5}$) and low-amplitude ($U=10^{-5}$) torques are shown at two different times $t_1=7500, t_2=30050$, where black arrows indicate time evolution. The corresponding theoretical torques are indicated by the blue and black solid lines, respectively. The vertical solid red line marks the eigenfrequency at $\omega=0.11136$.  }
    \label{fig:torque_peak}
\end{figure}

In Fig.~\ref{fig:torque_peak}, we show how the torque varies as a function of the forcing frequencies $\omega$. For the high-amplitude forcing  ($U=10^{-4}$, filled circles), prompt wave breaking leads to the travelling-wave torque $T_{\rm{tw}}$ (equation~\ref{ttw}). A power-law fit yields a $\omega$-dependence ($\sim \omega^{-0.85}$, red dashed line) very close the theoretical expectation ($\sim\omega^{-1}$). We attribute the small difference to two possible reasons: 1) we only have a very narrow frequency range; 2) if the gravity waves are not completely absorbed by the critical layer, a small fraction of reflection can cause the power to differ from -1.0.

For the low-amplitude forcing ($U=1 \times 10^{-5}$), the torque is shown as crosses ($t_1=7500$) and diamonds ($t_2=30050$) under the theoretical standing-wave torque $T_{\rm{sw}}$ curve (solid black line, equation~\ref{tsw}). The two moments are also shown in Fig.~\ref{fig:torque_amp1e5} with the corresponding central angular velocities labelled. Note the slight decrease in torque from the crosses to the diamonds, which is the result of spin-up and evolution away from resonance. This can be seen most clearly for $\omega=0.110$ and $0.112$, for which we indicate the evolution by the black arrows. The theoretical torques calculated from equation~(\ref{tsw}) are in quite good agreement with the values from simulations, although $20\%$ differences are observed (crosses).

For the intermediate-amplitude forcing ($U=3\times 10^{-5}$), we again show the torque at two different moments $t_1$ and $t_2$ (marked in Fig.~\ref{fig:torque_amp3e5} as diamonds and triangles). The black arrows indicate the evolution direction. Note that for $\omega=0.100$, $0.102$, $0.112$, $0.114$ and $0.116$, critical-layer formation has already occurred by the second moment, so that the triangles are in the travelling-wave torque regime. A linear fit is shown as the blue dashed line. Note the significant increase in the torque value for $\omega=0.100$, $0.102$ and $0.116$ (from diamonds to triangles).
For $\omega=0.104$, $0.106$, $0.108$ and $0.110$, the system is evolving away from resonance and we see a decrease in torque  from diamonds to triangles. For $\omega=0.100$, we also calculate the specific torque at $t_1=8350$ and $t_2=15260$ from the velocities obtained by solving equations (\ref{primitive1})--(\ref{primitive4}). These torque values (green circles) are in good agreement with the full Nek5000 simulations in Fig.~\ref{fig:torque_peak} (black diamonds and triangles).

\subsection{Changes to the phases and resonances of linear waves due to the mean flows}

The evolution of the system towards resonance and the gradual spinning-up of the core, seen in numerical simulations of subcritical waves, motivates us to study how the mean flow changes the resonance condition of the waves.

The local dispersion relation of ideal linear waves on an evolved background can be deduced from equations (\ref{linear1})--(\ref{linear3}) by taking a short-wavelength limit in which the operator $\mathcal{L}$ is replaced by multiplication by $k_r^2+m^2/r^2$, where $k_r(r)$ is the local radial wavenumber:
\begin{equation}
  \left(\Omega_\text{p}-\overline{\Omega}\right)\left[\left(\Omega_\text{p}-\overline{\Omega}\right)\left(k_r^2+\frac{m^2}{r^2}\right)-\frac{1}{r}\frac{\dd\overline{\zeta}}{\dd r}\right]=\frac{N^2}{r^2}.\label{ldr}
\end{equation}
This dispersion relation includes the effects of internal gravity waves (the $N^2$ term) and Rossby waves (the $\dd\overline{\zeta}/\dd r$ term), as well as the Doppler shift (the $\overline{\Omega}$ terms) due to the mean flow.

On the original background, which has $\overline{\Omega}=0$ and $N^2=C^2r^2$, the local dispersion relation reduces to
\begin{equation}
  k_r^2=k^2-\frac{m^2}{r^2},
\end{equation}
where $k=C/\Omega_\text{p}$ is the (constant) total wavenumber used in Section~\ref{s:gravity}. The wave propagates where $k_r^2>0$, i.e.\ for $r>r_\text{t}$, where $r_\text{t}=m/k$ is the radius of the turning point.\footnote{The wave is formally evanescent for $r<r_\text{t}$, although the behaviour of the Bessel function $J_m(kr)$ for very small $r$ is actually $\propto r^m$ rather than an exponential decay.} Consider a wave travelling inwards from the outer boundary $r=R$ towards the turning point and returning to the outer boundary. The total change in phase is given by
\begin{equation}
  \Delta\varphi=2\int_{r_\text{t}}^Rk_r\,\dd r+\frac{\pi}{2}, 
\label{shift}
\end{equation}
where we take $k_r$ to mean the positive square root of $k_r^2$, and the last term accounts for the phase change on reflection from the turning point (as can be deduced, for example, from the theory of the Airy function). For the original background, this integral can be evaluated analytically, giving
\begin{equation}
  \Delta\varphi=2m(\alpha-\arctan\alpha)+\frac{\pi}{2},
\end{equation}
where
\begin{equation}
  \alpha=\sqrt{\beta^2-1},\qquad
  \beta=\frac{X}{m}.
\end{equation}
Setting $\Delta\varphi=2\pi n$, where the positive integer $n$ is the radial mode number, gives an excellent approximation to the exact g~modes. In the case $m=2$, the phase-integral approximation gives $X=5.100,8.400,11.608,14.787,17.953$, etc., which compare very favourably with the exact values $X=5.136,8.417,11.620,14.796,17.960$, etc., found from the zeros of the Bessel function, especially for larger values of $n$.

When the background has been modified as a result of wave dissipation, this affects the dispersion relation of waves and therefore the value of the phase integral. The turning point moves and the local wavenumber is modified.

We find that the most important effect of the evolved background is the Doppler shift due to the mean flow, which reduces the wave frequency in the fluid frame. Neglecting the $\dd\overline{\zeta}/\dd r$ and $\dd\overline{b}/\dd r$ terms in equation~(\ref{ldr}), we simplify the dispersion relation to
\begin{equation}
  k_r^2\approx k^2\left(1-\frac{\overline{\Omega}}{\Omega_\text{p}}\right)^{-2}-\frac{m^2}{r^2}.
\end{equation}
The Doppler shift causes the radial wavenumber to increase (as noted in Section~\ref{s:linear}) and the turning point to move inwards, both of which increase the phase integral. Let the angular-velocity profile be
\begin{equation}
  \overline{\Omega}=\Omega_\text{m}f_\Omega(x),
\end{equation}
where $\Omega_\text{m}$ is the maximum angular velocity and $f_\Omega(x)$ is a dimensionless function with a maximum value of~$1$. Define the dimensionless spin parameter $\omega_\text{m}=\Omega_\text{m}/\Omega_\text{p}$, and assume that $0\le\omega_\text{m}<1$, so that no critical layer is present.

Working initially to first order in $\omega_\text{m}$, we find that the local change in the radial wavenumber is
\begin{equation}
  \delta k_r=\frac{k^2}{k_r}\frac{\overline{\Omega}}{\Omega_\text{p}}
\end{equation}
and the corresponding change in the phase integral is
\begin{equation}
  \delta\varphi=2\int_{r_\text{t}}^R\delta k_r\,\dd r.
\end{equation}
(To this order, the change in the location of the turning point does not need to be considered, since $k_r$ vanishes there.) Then
\begin{equation}
  \frac{\delta\varphi}{2\pi}\approx\frac{\omega_\text{m}}{\pi}\int_m^X\left(1-\frac{m^2}{x^2}\right)^{-1/2}f_\Omega(x)\,\dd x.
\end{equation}

If we now set $m=2$ and $X=20$ (typical of our numerical simulations) and take $f_\Omega$ to be the equilibrium angular-velocity profile $f_2$ defined in equation~(\ref{f2}), but modified as described there and renormalized to have a maximum of~$1$, i.e.
\begin{equation}
  f_\Omega(x)=\frac{f_2(x)-f_2(X)}{1-f_2(X)},
\label{fomega}
\end{equation}
then we find
\begin{equation}
  \frac{\delta\varphi}{2\pi}\approx1.66\,\omega_\text{m}.
\label{delta_phi_first_order}
\end{equation}
This result suggests that
only a partial spin-up of the central region is needed to change the phase of the wave by $2\pi$, which corresponds to the shift between neighbouring resonant peaks.

\begin{figure}
	\includegraphics[width=\columnwidth]{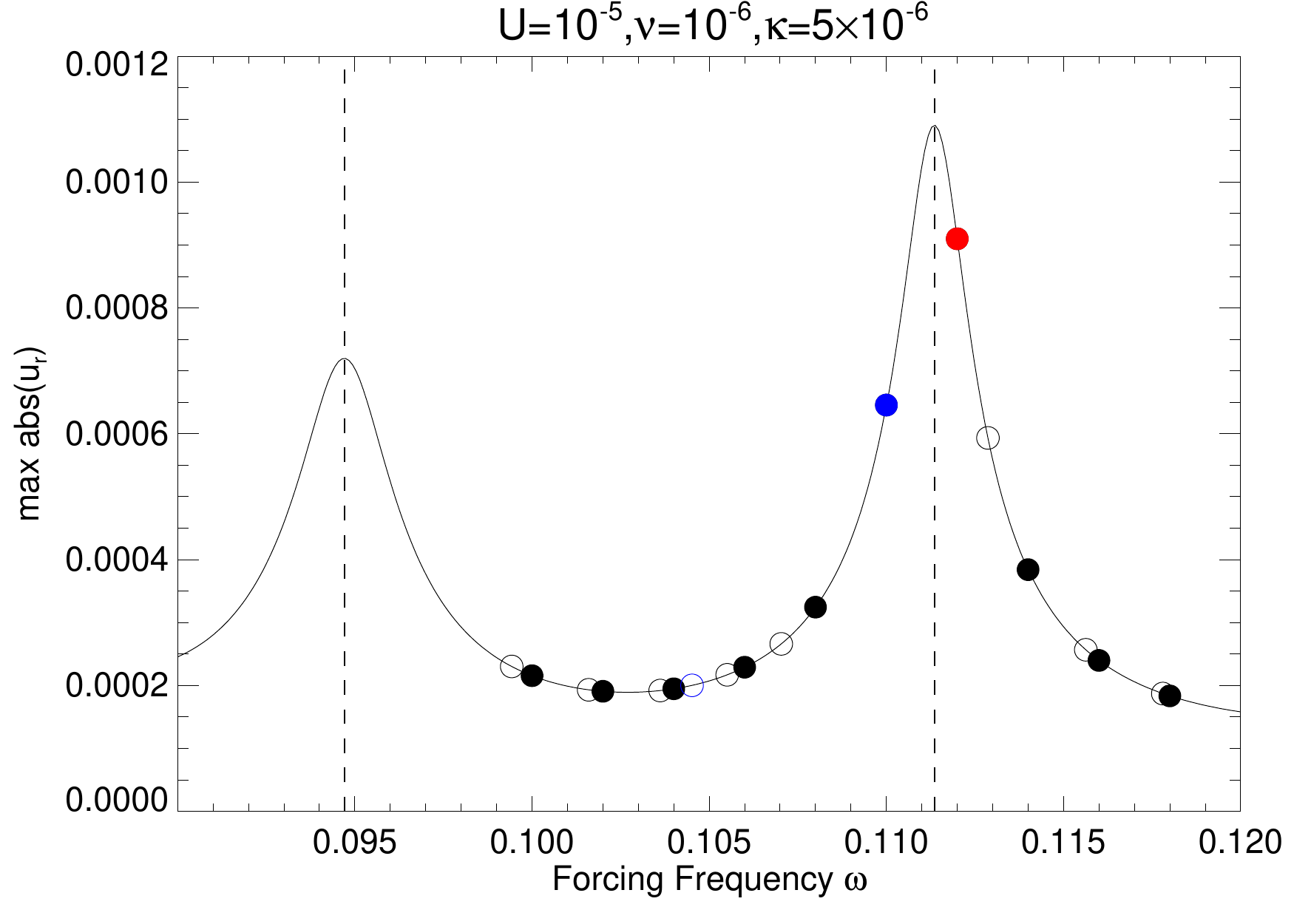}
    \caption{ Shifts of wave frequencies in the fluid frame due to the spin-up profile $\Omega(r)$ for the low-amplitude ($U=10^{-5}$) case, with $\nu=10^{-6}, \kappa=5\times 10^{-6}$). A phase shift of $\delta \phi=2\pi$ is translated to the frequency shift $\delta\omega=\omega_{\rm{eig2}}-\omega_{\rm{eig1}}$. The initial forcing frequencies are indicated by the filled circles, while the final, steady-state frequencies are shown as open circles. Note that the two small-detuning frequencies $\omega=0.112,0.110$ have very large shifts and are indicated by the red and blue symbols, respectively. }
    \label{fig:w_shift}
\end{figure}

In Fig.~\ref{fig:w_shift}, we illustrate the effect of the spin-up process by showing the original forcing frequencies (filled circles) and the equivalent forcing frequencies after the spin-up. In this low-amplitude forcing case, we calculate the steady-state angular velocity profile after spin-up (equation~\ref{omega_steady}) and its corresponding phase shift (equations \ref{ldr} and \ref{shift}). Then the phase shifts are transformed to shifts of the forcing frequency as if the background remained non-rotating, and we use open circles to indicate the equivalent forcing frequencies after spin-up. Note that the shifts are much larger if the forcing frequencies are close to resonances (e.g., the red and blue symbols). The red symbol actually shifts to such a low frequency that it is out of view.

\begin{figure}
	\includegraphics[width=\columnwidth]{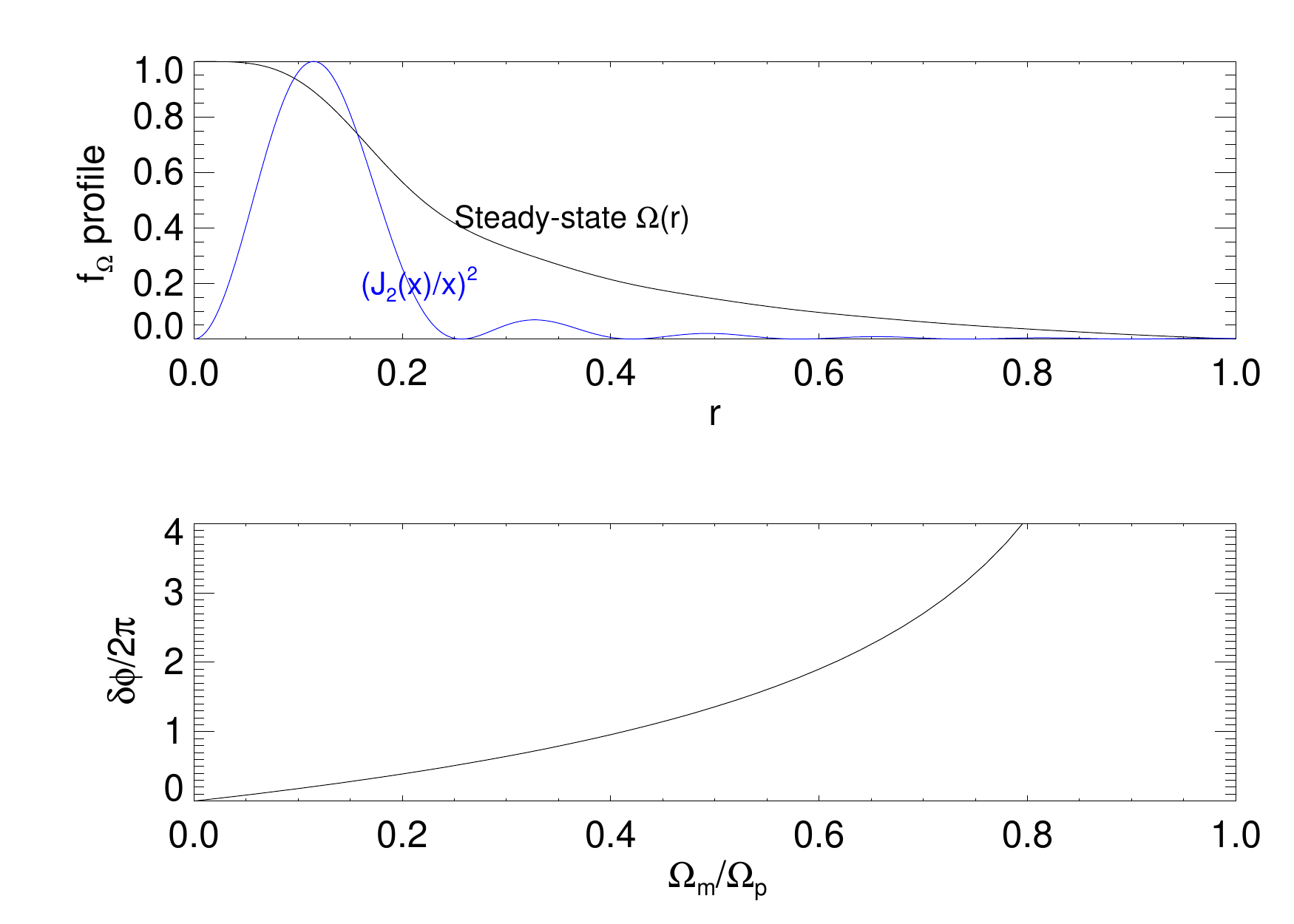}
    \caption{Top: Steady-state spin-up profile $\Omega(r)$ and angular-momentum deposition profile $(J_2(x)/x)^2$. Both are normalized to have a maximum of unity. Bottom: Phase change due to the steady-state spin-up profile, as a function of the dimensionless spin parameter $\omega_\text{m}=\Omega_\text{m}/\Omega_\text{p}$.}
    \label{fig:profile}
\end{figure}

The first-order approximation is valid only for small values of $\omega_\text{m}$. For greater degrees of spin-up, the phase shift is even larger than predicted by equation~(\ref{delta_phi_first_order}). A numerical evaluation (equations \ref{ldr} and \ref{shift}) of $\delta\varphi/2\pi$ as a function of $\omega_\text{m}$ is shown in Fig.~\ref{fig:profile}. This implies that a $\approx 42\%$ spin-up of the central region ($\omega_\text{m} \approx 0.42$) produces a large enough phase shift to move from one resonance to the next ($\delta\varphi = 2\pi$).

Let us now consider the changes in the phase integral due to the evolution of the vorticity and buoyancy profiles. The first-order contribution to the change in the radial wavenumber due to $\dd\overline{\zeta}/\dd r$ is
\begin{equation}
  \delta k_r\approx\frac{1}{2k_r}\frac{1}{\Omega_\text{p}}\frac{1}{r}\frac{\dd}{\dd r}\left[\frac{1}{r}\frac{\dd}{\dd r}\left(r^2\overline{\Omega}\right)\right]
\end{equation}
and the corresponding change in the phase integral is given by
\begin{align}
  \frac{\delta\varphi}{2\pi}&=\frac{\omega_\text{m}}{2\pi}\int_{m}^X\left(1-\frac{m^2}{x^2}\right)^{-1/2}\frac{1}{x}\frac{\dd}{\dd x}\left[\frac{1}{x}\frac{\dd}{\dd x}\left(x^2f_\Omega\right)\right]\,\dd x\nonumber\\
  &\approx-0.15\,\omega_\text{m},
\end{align}
where the numerical evaluation is again for $m=2$, $X=20$ and $f_\Omega$ as in equation~(\ref{fomega}). This calculation shows that the effect of the vorticity gradient is much smaller than that of the Doppler shift and in the opposite direction. Beyond the first-order approximation, 
the effect of the vorticity gradient is relatively unimportant for larger values of $\omega_\text{m}$.

The first-order contribution to the change in the radial wavenumber due to evolution of the buoyancy profile is
\begin{equation}
  \delta k_r\approx\frac{1}{2k_r}\frac{1}{\Omega_\text{p}^2}\frac{1}{r}\frac{\dd\overline{b}}{\dd r},
\end{equation}
which is harder to relate to the spin parameter $\omega_\text{m}$.
If we adopt the equilibrium buoyancy profile~(\ref{b_steady}), then we obtain
\begin{equation}
  \frac{\delta\varphi}{2\pi}\approx-\frac{0.2591|A|^2}{2\pi}\int_2^\infty\left(1-\frac{4}{x^2}\right)^{-1/2}f_1(x)\,\dd x\approx-0.13|A|^2,
\end{equation}
which is small for any subcritical wave. Hence the most important modification of the phase of the wave comes from the Doppler shift due to the spin-up of the central region.

\section{Implications for solar-type stars}
\label{s:implications}

\begin{figure}
	\includegraphics[width=\columnwidth,trim=5cm 0cm 4cm 0cm,clip=true]{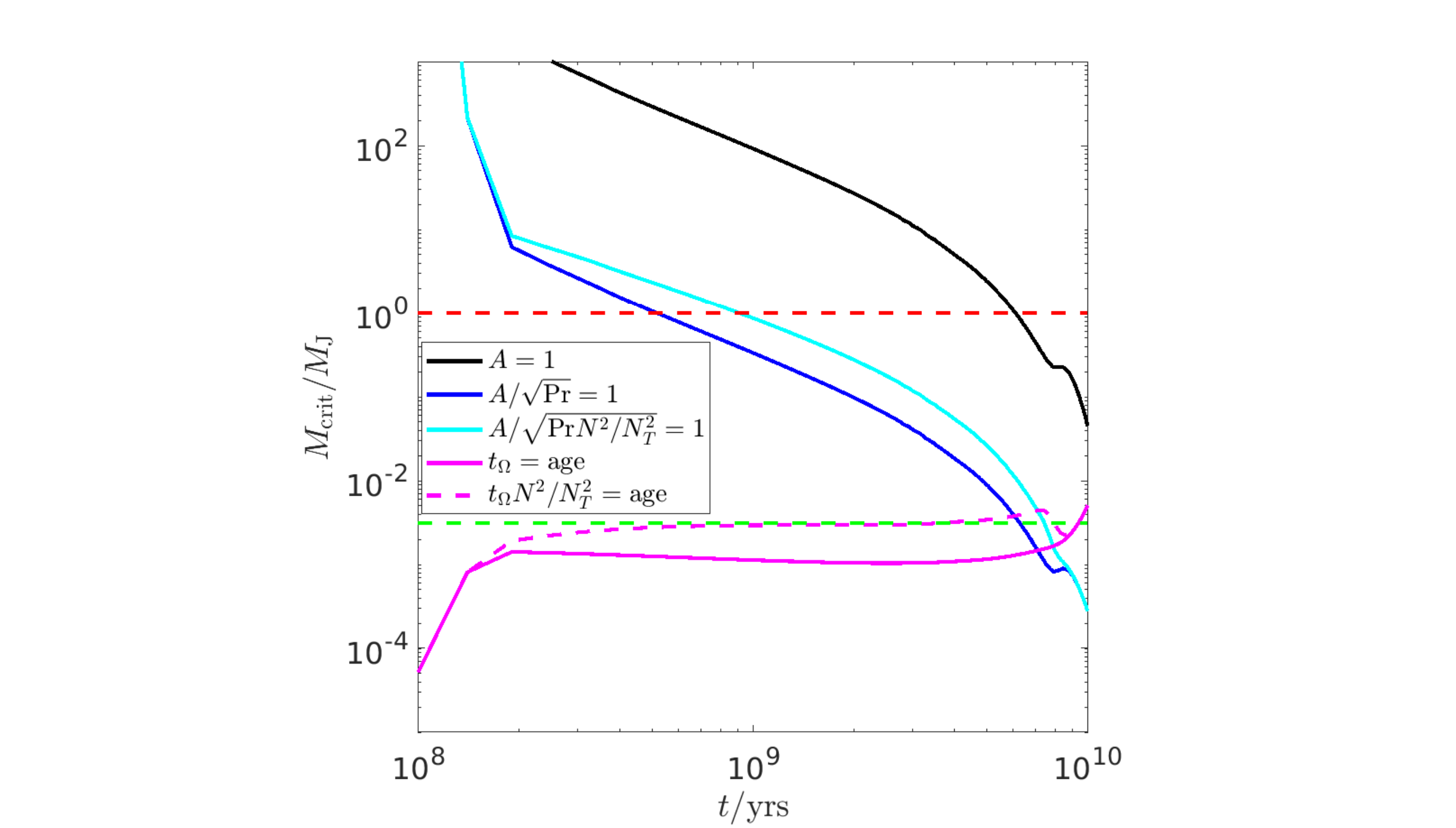}
    \caption{Critical mass for critical layer formation as a function of stellar age (see main text for a detailed explanation).}
    \label{fig:Mcrit}
\end{figure}


Our simulations and analysis build upon those of BO10, but they include the effect of resonances and the spin-up of the core by viscous and thermal damping of gravity waves. The implications for solar-type stars are as follows:

First, one of the most important conclusions of BO10 is that sufficiently massive exoplanets ($M \gtrsim 3M_\text{J}$ in the case of the present Sun, corresponding to wave amplitude $|A| \ge 1$) can induce gravity waves that break near the centre of a solar-type star and form an expanding critical layer that absorbs waves, leading to efficient tidal dissipation. However, this result applies only to off-resonance forcing, as the amplitude $|A|$ assumed by BO10 is independent of the forcing frequency. It is based on a travelling-wave solution \citep{Ogi07} and corresponds to a forcing frequency intermediate between adjacent resonances. If we consider the wave amplitude $A$ to be frequency-dependent as for a standing wave (Fig.~\ref{fig:A_w}), our high-amplitude forcing simulations (red line in Fig.~\ref{fig:A_w}) lead to immediate wave breaking for all forcing frequencies $\omega$, even though some of the off-resonance cases have $|A|$ as small as $0.3$--$0.5$, i.e.\ a factor of a few smaller than the critical amplitude ($|A|=1$) obtained in BO10. This may be because of the time-dependent nature of the response during the initial transient phase and the rapid spin-up due to wave damping. Furthermore, most of the simulations with intermediate-amplitude forcing (blue line in Fig.~\ref{fig:A_w}) still evolve to produce an expanding critical layer as a result of the gradual spin-up of the core due to  viscous and thermal wave damping. These cases correspond to $|A|$ as small as $\sim 0.1$. 

Thus, we can expect that the criterion for critical-layer formation should be lowered by at least a factor of a few (as large as $\sim 10$) for off-resonance forcing. 

Second, the very small values of the Prandtl number in stars open the possibility of critical-layer formation for significantly smaller forcing amplitudes, because of the slow viscous and thermal diffusive processes. As shown in equation~(\ref{omega_steady}) and Fig.~\ref{fig:steady_omega}, the equilibrium angular velocity  $\Omega_\text{eq}$ of the core depends on the factor $(1+1/\text{Pr})$ as well as the squared wave amplitude $|A|^2$.
On one hand, forcing frequencies close to resonance can result in a large $|A|$ and thus can lead to $\Omega_\text{eq}$ as large as the pattern speed $\Omega_\text{p}$ (Fig.~\ref{fig:steady_omega}, top and middle panels), which leads to the formation of a critical layer. On the other hand, since the solar core has $\text{Pr} \approx 2 \times 10^{-5}$, it is likely that $\Omega_\text{eq}$ can reach $\Omega_\text{p}$ even for small wave amplitudes $|A|$ (cf.~Fig.~\ref{fig:steady_omega}, bottom panel, although this is not for solar parameters and is based on the assumption that $\Omega\ll\Omega_\text{p}$.). Compensating for the $\text{Pr}$ dependence, we may lower the threshold wave amplitude for wave breaking to $|A| \gtrsim \sqrt{\text{Pr}} \approx 0.004$. This suggests that the critical mass of the orbiting exoplanet for critical-layer formation and wave absorption can be lowered to the order of a few Earth masses (since $A$ is proportional to the planetary mass), if this slow spin-up process is quick enough compared to other processes. 

If the source term (equation~\ref{s_omega_estimate}) is estimated and regarded as independent of time, then the timescale to produce a critical layer through spin-up due to thermal diffusion can be written as

\begin{equation}
  t_\Omega=\frac{\Omega_\text{p}}{\max S_\Omega}=\frac{30.88}{8|A|^2\kappa k^2}=\frac{5600\,\text{yr}}{|A|^2}\left(\frac{\kappa}{\kappa_\odot}\right)^{-1}\left(\frac{C}{C_\odot}\right)^{-2}\left(\frac{P}{\text{d}}\right)^{-2},
\label{t_omega}
\end{equation}
where we write the pattern speed $\Omega_\text{p}=2\pi/P$ in terms of an orbital period $P$, neglect $\nu$ compared to $\kappa$ and, for numerical evaluation, take the estimates $\kappa_\odot=1.65\times10^5\,\text{cm}^2\,\text{s}^{-1}$ and $C_\odot^2=7\times10^{-25}\,\text{cm}^{-2}\,\text{s}^{-2}$. This spin-up timescale $t_{\Omega}$ will be used to set the lower limit of the critical mass for critical-layer formation in Fig.~\ref{fig:Mcrit}. If the compositional contribution to buoyancy is included, $t_{\Omega}$ should be increased by a factor of $N^2/N_T^2 \sim 10$. This is because, as we argue in Appendix~\ref{s:appendix}, the effective thermal diffusivity is reduced by a factor of $N_T^2/N^2$. We note that the effective Prandtl number $(N^2/N_\text{T}^2)\,\text{Pr}$ is still very small for the Sun.


In Fig.~\ref{fig:Mcrit}, we show the critical mass $M_\text{crit}$ for wave breaking or critical-layer formation as a function of age, for a solar-mass star, in units of Jupiter's mass. Following BO10, the black line is based on a critical wave amplitude $A=1$ for the off-resonance forcing frequencies. Note that $M_\text{crit} \approx 3 M_\text{J}$ for the current Sun ($t=4.6$ Gyr). After taking into account the effect of small Prandtl number in the Sun, the $M_\text{crit}$ curve shifts downward to the blue line, based on equation~(\ref{omega_steady}). However, if the reduction of the effective thermal diffusivity due to the dominance of the non-thermal, compositional  contribution to the buoyancy is taken into account, the line changes to the cyan one. If we adopt $t_\Omega$ (equation~\ref{t_omega}) as the timescale to create a critical layer through the spin-up of the stellar core, then the planetary mass must be above the solid purple line (dashed purple line if the compositional buoyancy is included) in order for the critical layer to be formed within the age of the system. The dashed purple line provides the equivalent constraint taking into account the reduction of the effective thermal diffusivity. 
The red and green dashed horizontal lines correspond to the masses of Jupiter and the Earth, respectively. We tentatively conclude that planets above several Earth masses could lead to critical-layer formation and efficient tidal dissipation in a star similar to the Sun after several Gyr. At an early stage, the threshold is raised to about one Jupiter mass at around $1~\text{Gyr}$.

Our exploratory study has some important limitations. We considered only three forcing amplitudes and did not fully explore the effect of resonances on the critical wave amplitude for critical-layer formation. Our simulations show that the system can evolve through resonances in the case of low-amplitude forcing, while critical-layer formation occurs for intermediate and high-amplitude forcing. 

For large amplitude forcing ($U=10^{-4}$), higher $m$ modes are present as shown in BO10 (Figure 8 and 9), which are generated by wave breaking (or near critical layers). However, for the intermediate-amplitude forcing cases, we did not see obvious secondary wave generation from the weakly-nonlinear parametric instability. This is probably because, in our simulations, the dominant instability is the localized convective instability ($A>1$), which is stronger than the parametric instability ($A<1$). Higher $m$ daughter modes generated by non-linear mode-couplings like the parametric instability have relatively smaller growth rates. In fact, according to \citet{Bar11a}, the maximum growth rate of these unstable secondary waves is $\gamma_s \sim 0.02\omega_p \sim 0.002$, with a typical primary wave frequency $\omega_p\approx 0.1$. To overcome the damping, we require $\gamma_s \geq (1/2)(\nu+\kappa)k^2$. With typical values of $\nu \sim \kappa \sim 10^{-6}$, this gives $k \leq 45$, which translates to the constraints on the radial order $n$ and azimuthal wavenumber $m$: $n\leq 6$ and $m\leq 4$. For such mode couplings to operate, the daughter wave pair should also have a frequency detuning smaller than $\gamma_s$, which would be very difficult to achieve for the eigenfrequencies of the Bessel function $J_{m}$. To conclude, we find the secondary waves generated by parametric instability would be very difficult to observe in our simulations.

While our simulations are performed in 2D, a 3D simulation should be pursued further. We expect to find qualitatively similar results although some of the details may differ \citep{Bar11}. It would also be valuable to extend the quasi-linear approach to study the 3D problem.

We have neglected the role of magnetic fields. There are several possible effects that could be studied in future work. A large-scale poloidal magnetic field, if present in the core, would tend to suppress differential rotation and is therefore in competition with the spin-up process we have identified. It could also affect the propagation of the gravity waves themselves. Magnetic instabilities could play an important role: for example, the magnetorotational instability might lead to outward angular-momentum transport in the differentially rotating flow, or magnetic buoyancy  might expel flux from the core.

In addition to tidally forced waves, a broad spectrum of gravity waves can be generated by convection and propagate towards the centre, where they may deposit angular momentum and contribute to the evolution of the mean flow.

In real applications, the stellar eigenmodes are slowly changing owing to stellar evolution. The orbital evolution of the planet
can keep in pace with stellar evolution, leading to the resonance-locking scenario. In principle, we can slowly change the forcing frequency to mimic these effects. But, as we have shown, the nonlinear feedback of the waves on the background state can generate a differential rotation of the core that significantly changes the resonance conditions. The classical picture of resonance locking usually assumes solid-body rotation \citep[e.g.][]{Ma21}. Actually, only the central wavelength of the core needs to be spun up to significantly change the phase of the gravity waves and their resonances. This region is small (typically $\lambda \sim 0.01 R_\odot$)
and has a much smaller moment of inertia than the entire star, making it much more mobile. Thus the treatment of resonance locking in solar-type stars needs to be revised.

\section{Conclusions}
\label{s:conclusions}

In this study, we use a Boussinesq model (following BO10) to study the behaviour of tidally excited gravity waves in the radiative cores of solar-type stars. We study a circular 2D cavity using linear theory and nonlinear hydrodynamical simulations, to mimic the innermost regions of the stellar core. The boundary conditions are carefully chosen so that internal gravity waves (of azimuthal wavenumber $m=2$) are generated, allowing resonances with the eigenmodes and non-linear wave--mean-flow interactions to be studied.
We consider a range of forcing frequencies and three different forcing amplitudes (low, intermediate and high), as well as different values of the viscosity and thermal diffusivity, which are crucial for the wave dissipation.

Similar to the results of BO10, we find that waves break at the centre if the wave amplitude is sufficiently large. This occurs irrespective of the resonance condition in the case of high-amplitude forcing, and leads to the formation of a rotating core that acts as a critical layer 
that absorbs subsequent incoming waves. 

Building upon, and differently from, BO10, we pay particular attention to resonances and find that smaller wave amplitudes (as in the cases of intermediate and low-amplitude forcing) can still lead to the formation of a differentially rotating core through viscous and thermal wave dissipation and its feedback on the mean flow. The system can evolve towards or away from resonances with correspondingly increasing or decreasing wave amplitude. A critical layer can be formed without wave breaking, if the star has enough time to spin up the core to the pattern speed of the wave through thermal (or viscous) diffusion, and if the Prandtl number is sufficiently small. The formation of a critical layer and the subsequent absorption of incoming gravity waves leads to strong tidal dissipation with a smooth dependence on the forcing frequency, which has important implications for the evolution of close binary stars and short-period exoplanets.   Even if a critical layer is not formed, the small Prandtl number of the solar core means that the central region can be easily spun up to a significant fraction of the wave pattern speed. In this case the phase of the wave can be significantly altered, to the extent that the classical picture of resonance locking needs to be amended. 

We emphasize the success of the quasi-linear theory, developed in Section~\ref{s:analytical}, in describing the wave--mean-flow interactions, the development of differential rotation and the formation of critical layers. It is very promising since it can be applied to regimes that are unattainable in numerical simulations.

Other effects that can modify the differential rotation profile in the core should be studied further, in particular those due to a magnetic field.



\section*{Data Availability}
 The data generated in this research will be shared on reasonable request to the corresponding author.

\section*{Acknowledgements}

This project was initiated during the Kavli Summer Program In Astrophysics 2021: Fluid dynamics of the Sun and Stars. 
This work was supported by STFC grants ST/T00049X/1 (DAMTP, Cambridge),  ST/S000275/1 and and ST/W000873/1 (Leeds).



\bibliographystyle{mnras}
\bibliography{neksun} 




\appendix

\section{Wave propagation and damping with both thermal and compositional buoyancy}
\label{s:appendix}

If the Boussinesq model is extended to include two different sources of buoyancy (i.e.\ thermal and compositional), then the local dispersion relation of gravity waves in such a model becomes
\begin{equation}
  \omega+\ii\nu k^2=\frac{k_\text{h}^2}{k^2}\left(\frac{N_\text{t}^2}{\omega+\ii\kappa_\text{t}k^2}+\frac{N_\text{c}^2}{\omega+\ii\kappa_\text{c}k^2}\right),
\end{equation}
where  $k_\text{h}$ is the horizontal wavenumber, $k=\sqrt{k_r^2+k_\text{h}^2}$ is the total wavenumber, and the subscripts t and c refer to thermal and compositional. When diffusion is weak, the leading approximation to the dispersion relation is
\begin{equation}
  \omega^2\approx\frac{k_\text{h}^2}{k^2}N^2,
\end{equation}
where $N^2=N_\text{t}^2+N_\text{c}^2$ is the total squared buoyancy frequency. To first order in the diffusivities, waves of real $\omega$ and $k_\text{h}$ are radially attenuated according to
\begin{equation}
  \text{Im}(k_r^2)\approx-\ii\left(\nu+\frac{N_\text{t}^2}{N^2}\kappa_\text{t}+\frac{N_\text{c}^2}{N^2}\kappa_\text{c}\right)\frac{N^4}{\omega^5}k_\text{h}^4.
\end{equation}
The equivalent expression in the case of a single form of buoyancy would have simply $(\nu+\kappa)$ in the bracket.
In stars we typically have $\kappa_\text{t}\gg\nu,\kappa_\text{c}$. In this limit, unless the entropy gradient is extremely small, the damping of gravity waves is dominated by thermal diffusion but its effectiveness is reduced by a factor of $N_\text{t}^2/N^2$ relative to a calculation (such as our simulations in this paper) in which the buoyancy is assumed to be of purely thermal origin. To relate the stellar regime to the problem studied in this paper, we should identify $\kappa$ with $(N_\text{t}^2/N^2)\kappa_\text{t}$ and replace the actual Prandtl number $\text{Pr}\ll1$ with an effective Prandtl number $(N^2/N_\text{t}^2)\,\text{Pr}$, which is still very small for the Sun.





\bsp	
\label{lastpage}
\end{document}